\documentclass[twocolumn,prx,nofootinbib,superscriptaddress,preprintnumbers]{revtex4-1}

\usepackage{youngtab} 
\usepackage{amsmath,amsfonts,amssymb,bm,multirow}
\usepackage{dcolumn}
\usepackage{graphicx,float,xcolor,slashed, braket,float}
\usepackage[caption=false]{subfig}
\usepackage[T1]{fontenc} 

\definecolor{myblue}{rgb}{0.11,0.06,0.71}
\definecolor{myred}{rgb}{1.02,0,0.05}

\usepackage[
colorlinks=true, 
pdfstartview=FitV, 
linkcolor=blue, 
citecolor=myred,
urlcolor=blue, 
bookmarks=true,
bookmarksnumbered=true,
pdftitle={},
pdfauthor={}
]{hyperref}

\allowdisplaybreaks[1]

\newcommand{\be}{\begin{equation}}      
\newcommand{\ee}{\end{equation}}      
\newcommand{\bea}{\begin{eqnarray}}      
\newcommand{\eea}{\end{eqnarray}}
\newcommand{\im}{\mathrm{i}}            
\newcommand{\rmL}{\mathrm{L}}
\newcommand{\rmR}{\mathrm{R}}
\newcommand{\rmV}{\mathrm{V}}
\newcommand{\p}{\partial}
\newcommand{\ve}{\varepsilon}
\newcommand{\rmB}{\mathrm{B}}

\newcommand{\+}{\dagger}
\newcommand{\ra}{\rightarrow}

\newcommand{\Z}{{\mathbb Z}}

\newcommand{\diff}{\mathrm{d}}
\newcommand{\rmi}{\mathrm{i}}
\newcommand{\rme}{\mathrm{e}}

\newcommand{\rml}{\mathrm{L}}

\newcommand{\rmv}{\mathrm{V}}

\newcommand{\rmb}{\mathrm{B}}
\newcommand{\rmm}{\mathrm{M}}
\newcommand{\rmu}{\mathrm{U}}
\newcommand{\rmd}{\mathrm{D}}
\newcommand{\tr}{\mathrm{tr}}
\newcommand{\U}{\mathrm{U}}
\newcommand{\SU}{\mathrm{SU}}
\newcommand{\SO}{\mathrm{SO}}
\newcommand{\Sp}{\mathrm{Sp}}

\begin{document} 

\title{Finite-Density Massless Two-Color QCD at Isospin Roberge-Weiss Point and 't Hooft Anomaly}

\author{Takuya Furusawa}\email{furusawa@stat.phys.titech.ac.jp}
\affiliation{Tokyo Institute of Technology, Ookayama, Meguro, Tokyo 152-8551, Japan}
\affiliation{Condensed Matter Theory Laboratory, RIKEN, Wako, Saitama 351-0198, Japan}
\author{Yuya Tanizaki}\email{yuya.tanizaki@yukawa.kyoto-u.ac.jp}
\affiliation{Yukawa Institute for Theoretical Physics, Kyoto University, Kyoto 606-8502, Japan}
\author{Etsuko Itou}\email{itou@yukawa.kyoto-u.ac.jp}
\affiliation{Department of Physics and Research and Education Center for Natural Sciences, Keio University, 4-1-1 Hiyoshi, Yokohama, Kanagawa 223-8521, Japan}
\affiliation{Department of Mathematics and Physics, Kochi University, 2-5-1 Akebono-Cho, Kochi 780-8520, Japan}
\affiliation{Research Center for Nuclear Physics (RCNP), Osaka University, 10-1 Mihogaoka, Ibaraki, Osaka 567-0047, Japan}

\preprint{YITP-20-69}
\begin{abstract}
We study the phase diagram of two-flavor massless two-color QCD (QC$_2$D) under the presence of quark chemical potentials and imaginary isospin chemical potentials. 
At the special point of the imaginary isospin chemical potential, called the isospin Roberge--Weiss (RW) point, two-flavor QC$_2$D enjoys the $\mathbb{Z}_2$ center symmetry that acts both on quark flavors and the Polyakov loop. 
We find a $\mathbb{Z}_2$ 't~Hooft anomaly of this system, which involves the $\mathbb{Z}_2$ center symmetry, the baryon-number symmetry, and the isospin chiral symmetry. 
Anomaly matching, therefore, constrains the possible phase diagram at any temperatures and quark chemical potentials at the isospin RW point, and we compare it with previous results obtained by chiral effective field theory and lattice simulations. 
We also point out an interesting similarity of two-flavor massless QC$_2$D with $(2+1)$d quantum anti-ferromagnetic systems. 
\end{abstract}

\maketitle
\tableofcontents


\section{Introduction} \label{sec: intro}
Quantum Chromodynamics (QCD) is the fundamental theory of nuclear and hadron physics, and it provides various interesting phenomena in extreme conditions~\cite{Fukushima:2010bq}. 
At low temperatures and densities, the fundamental degrees of freedom, quarks and gluons, are confined inside color-singlet hadrons, and at high temperatures, they are liberated and form quark-gluon plasma (QGP). 
As the density increases, we have nucleon superfluidity, and at ultimately high densities, it is expected to transform into the superfluid phase of quark matter, called color-flavor locked (CFL) phase~\cite{Alford:2001dt, Alford:2007xm}. 

Since the system is strongly coupled in most regions of the QCD phase diagram, we should rely on the numerical lattice Monte Carlo simulation in order to obtain concrete understandings both qualitatively and quantitatively. 
When the number of colors $N_c$ satisfies $N_c\ge 3$, this Monte Carlo simulation is limited to the systems with zero-baryon densities, as the baryon chemical potential produces the sign problem~\cite{Barbour:1997ej, Muroya:2003qs}. 
Because of this issue, we still have no reliable first-principle computation for finite-density QCD, despite the fact that we are expecting rich dynamics and its importance inside neutron stars. 

An exception is $2$-color QCD, i.e., $N_c=2$, and we abbreviate it as QC$_2$D\footnote{Not to be confused with two-``dimensional'' QCD, often denoted as QCD$_2$.}. Because of the pseudo-reality of the $\SU(2)$ gauge group, the quark determinant can be shown to be real-valued even at finite baryon chemical potentials. 
Therefore, QC$_2$D with an even number of flavors, $N_f\in 2\mathbb{Z}$, does not suffer from the sign problem, and it is a good playground to test various ideas of finite-density QCD. 
QC$_2$D shares some important aspects of strong dynamics of $N_c=3$ QCD: low-energy excitations consist of color-singlet hadrons, and chiral symmetry is spontaneously broken at low temperatures. 
Because of this nature, finite-density QC$_2$D is gathering much attention both theoretically and numerically (see, e.g. Refs.~\cite{Nakamura:1984uz, Hands:1999md, Kogut:1999iv, Kogut:2000ek, Splittorff:2000mm, Muroya:2002ry, Nishida:2003uj, Kogut:2001na, Kogut:2003ju, Schafer:2002yy, Giudice:2004se, Metlitski:2005db, Alles:2006ea, Hands:2006ve, Hands:2007uc, Hands:2010gd, Cea:2006yd,  Lombardo:2008vc, Kanazawa:2009ks, Brauner:2009gu, Akemann:2010tv, Hands:2011ye, Cotter:2012mb, Kanazawa:2012zr, Kashiwa:2012xm, Boz:2013rca, Braguta:2016cpw, Leino:2017hgm, Bornyakov:2017txe, Astrakhantsev:2018uzd, Iida:2019rah, Boz:2019enj}). 
There are other options to obtain the sign-problem free setup in $N_c\ge 3$ QCD with nonzero chemical potentials, such as the imaginary chemical potential~\cite{Alford:1998sd, deForcrand:2002hgr, DElia:2002tig, DElia:2009pdy, deForcrand:2010he, Nagata:2011yf,Morita:2011eu, Bonati:2014kpa, Nagata:2014fra, Takahashi:2014rta} or the isospin chemical potential~\cite{Son:2000xc, Kogut:2002zg, Kouno:2011zu, Cea:2012ev, Brandt:2017oyy}, and they also acquire big interest. 
In this paper,  we study the phase structure of QC$_2$D combining these aspects from the viewpoint of symmetry. 

We have emphasized the similarity between QC$_2$D and $N_c\ge 3$ with the usefulness of $N_c=2$ for studying finite densities, but it is also important to know how $2$-color nuclear matter is different from that of our world. 
The most important difference is that there are color-singlet diquarks when $N_c=2$. The diquark has the baryon charge $1$, so the superfluid state of finite-density nuclear matter is basically described by Bose-Einstein condensation (BEC) of the diquark. 
When $N_c=3$, the baryons are fermions, the nuclear matter forms the Fermi surface of nucleons, and the superfluidity is caused by Bardeen-Cooper-Schrieffer (BCS) pairing of two nucleons and its condensation at low densities. 
Thus, the mechanism of low-density nuclear superfluidity in the $N_c = 2$ and $N_c = 3$ cases are different qualitatively, but the color-superconducting states at ultimately high densities behave in a similar manner between $N_c=2$ and $N_c\ge 3$. 
Another notable difference is the nature of chiral symmetry. For $N_c\ge 3$ QCD with $N_f$ massless flavors, flavor symmetry is $\SU(N_f)_\rmL\times \SU(N_f)_\rmR\times \U(1)$, and the chiral condensate breaks it down to $\SU(N_f)_\rmV\times \U(1)$ with $N_f^2-1$ massless Nambu-Goldstone (NG) bosons. 
When $N_c=2$, however, the flavor symmetry is enhanced to $\SU(2N_f)\supset \SU(N_f)_\rmL\times \SU(N_f)_\rmR\times \U(1)$ because of the pseudo-reality of $\SU(2)$ color~\cite{Smilga:1994tb, Peskin:1980gc}, which is sometimes referred to as Pauli-G\"{u}rsey symmetry. 
Therefore, chiral-symmetry breaking pattern is $\SU(2N_f)\to \Sp(N_f)$ with $2N_f^2-N_f-1$ NG bosons\footnote{We use the convention for the compact symplectic group, $\Sp(N)$, as $\Sp(1)\simeq \SU(2)$ in this paper. Another useful exceptional isomorphism is $\Sp(2)\simeq \mathrm{Spin}(5)$. For details, see Appendix~\ref{sec:symplectic}.
We note that, in other literature, people sometimes use a different convention, $\Sp(2)=\SU(2)$, so there are factor $2$ difference of the arguments between  these conventions.}, which consist of $N_f^2-1$ mesons and $N_f^2-N_f$ diquarks. 
Importantly, with small current quark mass, the lightest diquarks have the same mass with pions. 
This clearly tells us why Monte Carlo simulations of QC$_2$D do not encounter the fake early onset of baryon-number densities, which is currently one of the biggest obstacles to tackle the low-temperature nuclear matter for $N_c= 3$~\cite{Barbour:1997ej, Cohen:2003kd, Nagata:2012tc, Tanizaki:2015rda}. 

The purpose of this paper is to explore the rigorous nature of the QC$_2$D phase diagram with exactly massless quarks. 
Even when the numerical lattice simulation is free from the sign problem, the chiral extrapolation requires careful, systematic studies with a lot of computational costs. 
Moreover, the chiral symmetry of the continuum theory suffers from lattice discretization, so we must have good knowledge about the nature of the chiral limit in order to understand such numerical simulations.  

It is, however, usually quite difficult to give a rigorous constraint on the phase diagram. 
It is widely thought that finite-temperature quantum field theories are mapped to classical statistical systems in the one-lower-dimensional space.  
Since classical phases of matters are classified by Ginzburg--Landau paradigm, such descriptions may accept any kind of phase transitions depending on the effective coupling constants. 
This expectation is correct in many systems, but a crucial exception was found for pure Yang-Mills theories~\cite{Gaiotto:2017yup}. 
Thanks to the existence of the center symmetry, thermal Yang-Mills theory behaves as if it is a quantum matter even at high temperatures, and, surprisingly, their phases are constrained by an 't~Hooft anomaly: Anomaly matching requires that the center symmetry or $CP$ symmetry at $\theta=\pi$ has to be broken at any temperatures.  
Motivated by this discovery, recent studies elucidate that QCD also enjoys the similar constraint even though it does not have a good center symmetry~\cite{Shimizu:2017asf, Tanizaki:2017qhf, Tanizaki:2017mtm, Dunne:2018hog} (For other related developments, see, e.g., Refs.~\cite{Witten:2016cio, Tachikawa:2016cha, Tanizaki:2017bam, Kikuchi:2017pcp, Komargodski:2017dmc, Komargodski:2017smk,  Wang:2017loc,Gaiotto:2017tne,  Yamazaki:2017dra, Guo:2017xex,  Sulejmanpasic:2018upi, Tanizaki:2018xto,  Kobayashi:2018yuk,  Tanizaki:2018wtg,  Anber:2018jdf,   Anber:2018xek,Cordova:2018acb, Armoni:2018bga, Yonekura:2019vyz,  Karasik:2019bxn,  Nishimura:2019umw,Cordova:2019jnf,Cordova:2019uob,  Misumi:2019dwq, Cherman:2019hbq, Bolognesi:2019fej, Tanizaki:2019rbk, Furusawa:2020kro, Sulejmanpasic:2020zfs}). 
Due to the absence of center symmetry, we have to choose a specific boundary condition of matter fields in order to obtain a nontrivial constraint, and this is equivalent to introduce imaginary chemical potentials. 
Thus, the phase diagram of QC$_2$D also has a chance to get a severe constraint under the presence of nonzero imaginary chemical potentials. 

In this paper, we study the phase diagram of massless two-flavor QC$_{2}$D in detail at finite quark chemical potentials and imaginary isospin chemical potentials. 
This setup does not suffer from the sign problem, so our predictions can be confirmed by lattice Monte Carlo simulations. 
At the special value of the imaginary isospin chemical potential, $\theta_I=-\im \mu_I L=\pi/2$, this theory has a $\mathbb{Z}_2$ symmetry acting on the quark flavor and Polyakov loop at the same time, which can be thought of as the center symmetry. 
We refer this special point as the isospin Roberge--Weiss (RW) point.
We find the $\mathbb{Z}_2$ anomaly related to this center symmetry and also chiral symmetry, and we discuss its implications to the finite-density phase diagram at the isospin RW point. 

\begin{figure*}[t] 
\includegraphics[width=15cm]{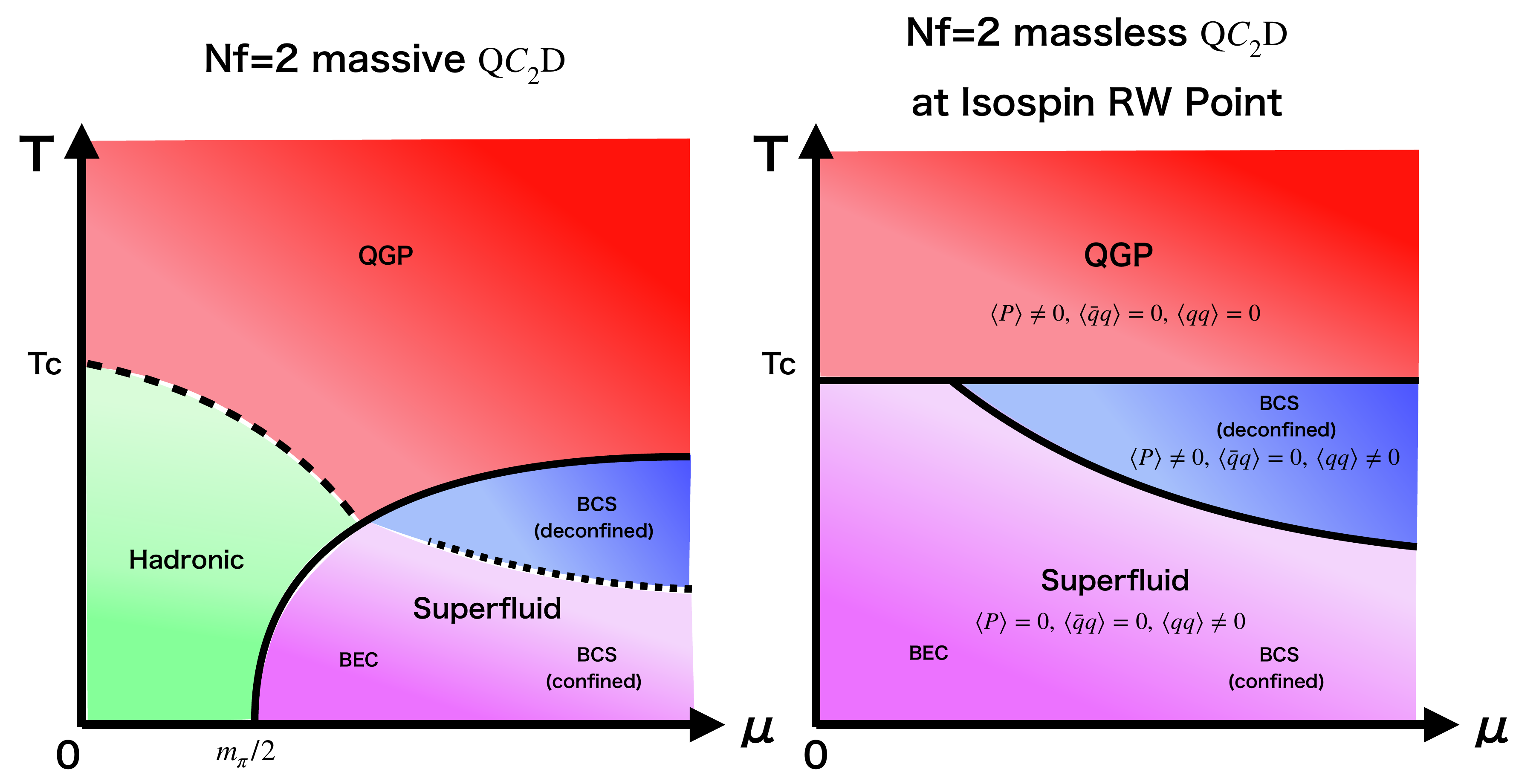}
\caption{(Left)~A schematic phase diagram of two-flavor massive QC$_2$D with the thermal boundary condition, predicted by theoretical and numerical works. 
Since this theory has neither the center nor chiral symmetry, the boundaries of phases expressed by dashed curves will be a crossover. (Right)~A possible phase diagram for two-flavor massless QC$_2$D at isospin RW point, which we give in this work. Here, we  discuss the spontaneous symmetry breaking of the center, chiral and baryon-number symmetries, and  classify the phases. Each boundary of phases is expected to be a critical line associated with a second-order phase transition.}
\label{fig:phase-diagram}
\end{figure*} 

Let us summarize our main result by comparing the phase diagrams of QC$_2$D with the thermal boundary condition and QC$_2$D at the isospin RW point. 
The left panel of Fig.~\ref{fig:phase-diagram} shows the schematic phase diagram of two-flavor QC$_2$D with \textit{massive} quarks with the thermal boundary condition, which seems to be consistent with all data given by the recent lattice simulations~\cite{Hands:2006ve, Braguta:2016cpw, Bornyakov:2017txe, Astrakhantsev:2018uzd, Iida:2019rah, Boz:2019enj}. 
However, we should note that the definition of phases are rather ambiguous because we have no exact order parameters for the confinement nor for chiral symmetry breaking, so different conventions are used between those papers. Accordingly, the dashed curves in the left panel of Fig.~\ref{fig:phase-diagram} can be the crossover lines.

In the right panel of Fig.~\ref{fig:phase-diagram}, we show one of the possible scenarios for the phase diagram of two-flavor QC$_2$D with \textit{massless} quarks at the isospin RW point. 
In this setup, as we shall reveal in this paper, we can discuss the spontaneous breaking of the center, chiral and baryon-number symmetries, and the corresponding order parameters are given by the Polyakov loop $\langle P \rangle$, the chiral condensate $\langle \bar{q} q \rangle$ and the  diquark condensate $\langle q q \rangle$, respectively.
We find the mixed anomaly between these three symmetries, and we conclude that, at least, one of these symmetries must be spontaneously broken at any temperatures and chemical potentials, in order to satisfy the anomaly matching condition. 
Figure~\ref{fig:phase-diagram} satisfies this requirement, and we shall discuss it in more details in Sec.~\ref{sec:phase_diagram}.

This paper is organized as follows. In Sec.~\ref{sec:symmetry}, we give a brief overview of QC$_2$D. We discuss global symmetry, especially emphasizing details on the discrete symmetry in this section. 
In Sec.~\ref{sec:perturbative_anomaly}, we discuss the anomaly matching condition on the perturbative triangle anomaly. 
This perturbative anomaly is useful to understand the phases at $T=0$. 
In Sec.~\ref{sec:discrete_anomaly}, we discuss the more subtle anomaly, i.e., discrete anomaly, related to the $\mathbb{Z}_2$ center symmetry at the isospin RW point.  
In Sec.~\ref{sec:phase_diagram}, we shall see how the discrete anomaly constrains the possible phase diagram of massless  QC$_2$D at the isospin RW point. 
In Sec.~\ref{sec:analogy}, we point out that the discrete anomaly of QC$_2$D at isospin RW point has a similar structure to that of $(2+1)$d quantum anti-ferromagnet. 
In Sec.~\ref{sec:summary}, we summarize the results. 
We describe our convention on Euclidean Dirac and Weyl spinors in Appendix~\ref{sec:gamma_matrix}, and useful facts on the compact symplectic group, $\Sp(N)$, in Appendix~\ref{sec:symplectic}.

\section{Global Symmetries of Massless QC$_2$D}\label{sec:symmetry}

This section is devoted to an overview of QC$_2$D related to its global symmetry. 
We first give a review on the global symmetry of massless QC$_2$D with $N_f$ Dirac flavors, paying attention to incidental details on the global structure. 
We discuss how the symmetry is affected by quark chemical potentials, $\mu$, and also by isospin chemical potentials, $\mu_I$, for an even number of flavors, $N_f\in 2\mathbb{Z}$. 
We note that QC$_2$D suffers from the sign problem if both $\mu$ and $\mu_I$ are present, but we emphasize that this does not occur if we introduce the imaginary isospin chemical potential, $\mu_I=\im \theta_I/L$, instead. Thanks to the RW periodicity for $\theta_I$, we shall find the $\mathbb{Z}_2$ center symmetry at the isospin version of RW point, $\theta_I=\pi/2$. 
In particular, for $N_f=2$, we concretely discuss the property of Nambu-Goldstone bosons as perturbations from the vacuum at $\mu=\mu_I=0$. 

\subsection{$\SU(2N_f)/\mathbb{Z}_2$ Chiral Symmetry}\label{sec:Pauli_Gursey}

We consider QC$_2$D with massless quarks, $\psi_{\mathrm{D},i}$, $i=1,\ldots, N_f$. The quark kinetic term is given by the Dirac Lagrangian,
\be
\sum_{i=1}^{N_f}\overline{\psi}_{\mathrm{D},i} \gamma^\nu (\p_\nu+\im a_\nu) \psi_{\mathrm{D},i}, 
\ee
where $a=a_\nu \diff x^\nu$ is the $\SU(2)$ gauge field\footnote{Throughout this paper, we denote the dynamical $\SU(2)$ gauge fields with the lower case, $a_\mu$, which is locally a Hermitian $2\times 2$ matrix-valued field. Upper case letters, $A,B,\ldots$, are reserved for background gauge fields of global symmetries. }. 
Throughout this paper, when we use Dirac spinors, we put the subscript ``$\mathrm{D}$'' because we mainly use Weyl spinors in this paper. Convention of Euclidean spinors used in this paper is summarized in Appendix~\ref{sec:gamma_matrix}. 

The motivation to use Weyl spinors is the existence of larger chiral symmetry. 
It is known that QC$_2$D has chiral symmetry $\SU(2N_f)/\mathbb{Z}_2$, which is larger than the usual one, $[\SU(N_f)_\rmL\times \SU(N_f)_\rmR\times \U(1)_\rmV]/(\mathbb{Z}_{N_f}\times \mathbb{Z}_2)$~\cite{Smilga:1994tb, Peskin:1980gc}, and it is easier to observe this symmetry in the Weyl notation. 
This enlarged symmetry appears due to the pseudo-real nature of the $\SU(2)$ gauge group. 
In order to see it, we decompose the $4$-component Dirac spinors into $2$-component Weyl spinors as follows: 
\be
\begin{split}
\psi_\mathrm{D}&=\begin{pmatrix}
\psi\\
(\ve_{\mathrm{color}}\otimes \ve_{\mathrm{spin}}) \bar{\tilde{\psi}}
\end{pmatrix}, 
\\
\overline{\psi}_\mathrm{D}&=\begin{pmatrix}
\tilde{\psi}(\ve_{\mathrm{color}}\otimes \ve_{\mathrm{spin}})^T \,\,, & \bar{\psi}
\end{pmatrix}. 
\end{split}
\label{eq:Dirac_Weyl}
\ee
Here, $\ve_{\mathrm{spin}}$'s denote the invariant tensors of the spacetime symmetry, $\mathrm{Spin}(4)\simeq \SU(2)\times \SU(2)$, and $\ve_{\mathrm{color}}$ is the invariant tensor of $\SU(2)$ color. 
For details, see the discussion around (\ref{eq:Dirac_Weyl_appendix}) of Appendix~\ref{sec:gamma_matrix}. An important point is that $\psi$ and $\tilde{\psi}$ are left-handed Weyl spinors, and both of them are in the defining representation, $\bm{2}$, of $\SU(2)$. 
The Dirac Lagrangian now becomes 
\be
\begin{split}
\sum_{i=1}^{N_f}&\overline{\psi}_{\mathrm{D},i} \gamma^\nu (\p_\nu+\im a_\nu) \psi_{\mathrm{D},i}
\\
&= \sum_{i=1}^{N_f} \left(\bar{\psi}_i \overline{\sigma}^\nu(\p_\nu+\im a_\nu) \psi_i+\bar{\tilde{\psi}}_i \overline{\sigma}^\nu (\p_\nu+\im a_\nu)\tilde{\psi}_i\right), 
\end{split}
\ee
up to integration by parts. 
The common chiral symmetry, $\SU(N_f)_\rmL\times \SU(N_f)_\rmR$, rotates $\psi$ and $\tilde{\psi}$, separately, but it is now evident that we can mix $\psi$ and $\tilde{\psi}$ as they share the same Lorentz and gauge structures. 
In other words, one can rewrite the quark kinetic term as
\be
\overline{\Psi}\overline{\sigma}^\nu D_\nu \Psi, 
\ee
where $D=\diff+\im a$ is the covariant derivative, and $\Psi$ represents $2N_f$ left-handed Weyl fermions defined as
\be
\Psi=\begin{pmatrix}
\psi_i\\
\tilde{\psi}_j
\end{pmatrix}.
\ee
At the classical level, this Lagrangian enjoys $\U(2N_f)$ symmetry, 
\be
\Psi\mapsto U\Psi,
\quad U\in \U(2N_f). 
\ee
Because of the Adler--Bell--Jackiw (ABJ) anomaly~\cite{Adler:1969gk, Bell:1969ts}, the fermion integration measure is not invariant under the continuous Abelian part of $\U(2N_f)$, so the theory is invariant only under $\SU(2N_f)$ symmetry\footnote{Let us comment on the discrete part. Writing $\U(2N_f)=[\SU(2N_f)\times \U(1)]/\mathbb{Z}_{2N_f}$, this $\U(1)$ symmetry is explicitly broken by ABJ anomaly as $\U(1)\to \mathbb{Z}_{2N_f}$. This discrete subgroup is identified as the center of $\SU(2N_f)$, so the $\mathbb{Z}_{2N_f}$ quotient gives $[\SU(2N_f)\times \mathbb{Z}_{2N_f}]/\mathbb{Z}_{2N_f}\simeq \SU(2N_f)$. }. 
This enhanced flavor symmetry is sometimes referred to as the Pauli-G\"{u}rsey symmetry. 

We note that the center element of $\SU(2)$ color group gives the gauge identification, $\Psi\sim -\Psi$, so the fermion parity can be eaten by gauge redundancy. In other words, any color-singlet operators are bosonic in this theory. 
As a consequence, $\mathbb{Z}_2=\{\pm \bm{1}_{2N_f}\}\subset \SU(2N_f)$ does not act on any gauge-invariant local operators. 
Therefore, the actual symmetry should be regarded as the quotient $\SU(2N_f)/\mathbb{Z}_2$. 

Let us identify how the common chiral symmetry, $[\SU(N_f)_\rmL\times \SU(N_f)_\rmR \times \U(1)_\rmV]/\mathbb{Z}_{N_f}$, is embedded into $\SU(2N_f)$. 
For this purpose, we note that, in the Dirac notation, $\psi_D$ transforms as the defining representations under the chiral symmetry, while $\overline{\psi}_D$ belongs to the conjugate representations. 
Therefore, we can summarize the charges of left-handed spinors, $\psi$ and $\tilde{\psi}$, under the $\SU(2)$ gauge and the ordinary chiral symmetries as follows:
\be
\begin{array}{c||c||c|c|c}
	& \SU(2) & \SU(N_f)_\rmL & \SU(N_f)_\rmR & \U(1)_\rmV \\ \hline
\psi & \bm{2} & \bm{N}_f & \bm{1} & 1\\
\tilde{\psi} & \bm{2} & \bm{1} & \overline{\bm{N}}_f & -1
\end{array}
\label{eq:charge_table}
\ee
Therefore, $(V_\rmL, V_\rmR, \rme^{\im \alpha})\in \SU(N_f)_\rmL\times \SU(N_f)_\rmR\times U(1)_\rmV$ is embedded into $\SU(2N_f)$ in the following fashion: 
\be
\begin{pmatrix}
\rme^{\im \alpha} V_\rmL & \bm{0}\\
\bm{0} & \rme^{-\im \alpha} (V_\rmR)^*
\end{pmatrix}\in \SU(2N_f). 
\label{eq:embedding_chiral_symmetry}
\ee
Let us put $\omega=\rme^{2\pi\im/(2N_f)}$. 
Then, $(V_\rmL,V_\rmR,\rme^{\im \alpha})$ and $(\omega V_\rmL, \omega V_\rmR, \omega^{-1}\rme^{\im \alpha})$ give the same element of $\SU(2N_f)$.
Thus, we identify the subgroup as 
\be
\frac{\SU(N_f)_\rmL\times \SU(N_f)_\rmR \times \U(1)_\rmV}{\mathbb{Z}_{N_f} \times \mathbb{Z}_2 }\subset {\SU(2N_f)\over \mathbb{Z}_2}. 
\label{eq:chiral_symmetries_relation}
\ee
Here, we take the $\mathbb{Z}_2$ quotient of both sides to take into account the gauge identification. 

Lastly, let us give a review on the chiral symmetry breaking of this theory in the vacuum~\cite{Kogut:1999iv, Kogut:2000ek}. 
When $N_f$ is not too large, the confining force is sufficiently strong so that we have chiral symmetry breaking. 
As an order parameter of chiral symmetry breaking, the most natural operator is the quark bilinear operator, which is spin singlet and color singlet. In our case, it is given by 
\be
\Sigma_{ij}= \Psi_i \Psi_j \equiv (\ve_{\mathrm{color}})^{c_1 c_2} (\ve_{\mathrm{spin}})^{\alpha \beta} \Psi_{\alpha, c_1, i} \Psi_{\beta, c_2, j}, 
\ee
where $c_{1,2}\in \{1,2\}$ are $\SU(2)$ color indices, $\alpha,\beta\in \{1,2\}$ are spin indices, and $i,j$ are $\SU(2N_f)$ flavor indices. 
In order to make this operator color and spin singlet, the fermionic wave function must be anti-symmetric under both color and spin. Because of Fermi statistics, this fermion bilinear is anti-symmetric under the $\SU(2N_f)$ flavor label, too:
\be
\Sigma^T=-\Sigma. 
\ee
Now, assume that chiral condensate is as symmetric as possible away from the origin.
Then, such an example of vacuum expectation values is given by 
\be
\bigl\langle\Sigma \bigr\rangle\propto \Sigma_0 = 
\begin{pmatrix}
\bm{0} & \bm{1}_N\\
-\bm{1}_N & \bm{0}
\end{pmatrix}. 
\ee
This causes spontaneous chiral symmetry breaking as (for details, see Appendix~\ref{sec:symplectic})
\be
\frac{\SU(2N_f)}{\mathbb{Z}_2 } \to \frac{\Sp(N_f)}{\mathbb{Z}_2}. 
\label{eq:chiral_SSB_vacuum_general}
\ee
The broken symmetry forms the coset $\SU(2N_f)/\Sp(N_f)$ describing $2N_f^2-N_f-1$ massless NG bosons.
They consist of $N_f^2-1$ massless mesons, $\tilde{\psi}\psi\sim \overline{\psi}_{\mathrm{D},\rmR}\psi_{\mathrm{D},\rmL}$, associated with the common pattern of chiral symmetry breaking, $\SU(N_f)_\rmL\times \SU(N_f)_\rmR\to \SU(N_f)_\rmV$, and also of $N_f^2-N_f$ massless diquarks, $\psi\psi$ and $\tilde{\psi}\tilde{\psi}$, which appear as NG bosons due to the enhanced chiral symmetry. 
Explicit form of the chiral effective Lagrangian for $N_f=2$ shall be discussed later in Sec.~\ref{sec:chiral_EFT_Nf2}. 

\subsection{Quark and Isospin Chemical Potentials}

In this subsection, we consider the case the Dirac flavors $N_f$ is even. 
We first introduce the quark chemical potential, $\mu$, and discuss how the global symmetry is affected. 
After that, we further introduce the isospin chemical potential, $\mu_I$. 
We also comment on the sign problem when we introduce both quark and isospin chemical potentials. 
We show that we can evade the sign problem when we consider $\mu\in\mathbb{R}$ and $\mu_I\in \im\, \mathbb{R}$.

In order to discuss the effect of chemical potentials to $\SU(2N_f)$ chiral symmetry, we need to identify the generators $T_Q$ and $T_I$ of quark-number and isospin symmetries, respectively, in the Weyl representation $\Psi$, and then we add the chemical potential terms, 
\be
-\overline{\Psi}\overline{\sigma}^0 (\mu T_Q+\mu_I T_I)\Psi. 
\ee
We should find the subgroup of $\SU(2N_f)$ that preserves these terms. 

The quark number symmetry is nothing but $\U(1)_\rmV$ of $[\SU(N_f)_\rmL\times \SU(N_f)_\rmR\times U(1)_\rmV]/\mathbb{Z}_{N_f}$. 
Using the embedding (\ref{eq:embedding_chiral_symmetry}) into $\SU(2N_f)$, we obtain 
\be
T_Q=\begin{pmatrix}
\bm{1}_{N_f} & \bm{0}\\
\bm{0} & -\bm{1}_{N_f}
\end{pmatrix}. 
\ee
In order to preserve the chemical potential term, $U\in \SU(2N_f)$ must satisfy $U^\dagger T_Q U = T_Q$. 
It is easy to see that this is true if and only if 
\be
U\in \frac{\SU(N_f)_\rmL\times \SU(N_f)_\rmR\times U(1)_\rmV}{\mathbb{Z}_{N_f}}. 
\ee
Thus, we obtain the ordinary chiral symmetry of QCD. 
Its physical interpretation is very clear. The enlarged chiral symmetry $\SU(2N_f)$ exists because we can rotate the quark and anti-quark, $\psi$ and $\tilde{\psi}$, but the chemical potential introduces the unbalance between them. 
Therefore, we can no longer rotate between $\psi$ and $\tilde{\psi}$. 

Next, let us consider about the isospin chemical potential for even $N_f$. 
We then classify those $N_f$ quarks into $N_f/2$ generations of the $\SU(2)$ doublets, and call those doublets as the up and down sectors,
\be
\psi=\begin{pmatrix}
u\\
d
\end{pmatrix},\; 
\tilde{\psi}=\begin{pmatrix}
\tilde{u}\\
\tilde{d}
\end{pmatrix}. 
\ee
We have thus identified the isospin symmetry as $\SU(2)\simeq \SU(2)\otimes \bm{1}_{N/2}\subset \SU(N_f)_\rmV$, and we denote their generators (Pauli matrices) as $\tau_1,\tau_2,\tau_3$. 
Since $\psi$ is in the fundamental representation while  $\tilde{\psi}$ is in the anti-fundamental representation as we have seen in (\ref{eq:embedding_chiral_symmetry}), its Cartan subgroup is generated by 
\be
\begin{split}
T_I&=\begin{pmatrix}
\tau_3\otimes \bm{1}_{N_f/2} &  \\
  & -\tau_3\otimes \bm{1}_{N_f/2}
\end{pmatrix}
\\
&=\begin{pmatrix}
\bm{1}_{N_f/2} & \bm{0} & & \\
\bm{0} & -\bm{1}_{N_f/2} & & \\
& & -\bm{1}_{N_f/2} & \bm{0} \\
& & \bm{0} & \bm{1}_{N_f/2}
\end{pmatrix}. 
\end{split}
\ee
The isospin chemical potential introduces the unbalance between $u$ and $d$ quarks, so we loose the rotation between them. 
However, we still have the symmetry that rotates between $u$ and $\tilde{d}$, so the remnant symmetry is isomorphic to the ordinary chiral symmetry $[\SU(N_f)\times \SU(N_f)\times U(1)]/\mathbb{Z}_{N_f}$, although they act differently on a given basis $\Psi=(\psi, \tilde{\psi})^T$. 
This isomorphism between the stabilizer groups of $T_Q$ and of $T_I$ comes from the fact that an element $S\in \SU(2N_f)$ relates $T_I$ and $T_Q$ as follows:
\be
S^\dagger T_I S = T_Q. 
\ee
Explicit form of $S$ exchanges $d$ and $\tilde{d}$, while it leaves $u$ and $\tilde{u}$. 
At the massless point, the quark chemical potential and isospin chemical potential can be interchanged by a chiral rotation in $\SU(2N_f)$. 
Therefore, these two chemical potentials are equivalent physically at the massless point. 

When we introduce both chemical potentials, the symmetry must preserve 
\be
\mu T_Q+\mu_I T_I .
\ee
Because of the quark chemical potential, chiral symmetry is explicitly broken as 
\be
\SU(2N_f) \xrightarrow{\,\, \mu\,\, } \frac{\SU(N_f)_\rmL \times \SU(N_f)_\rmR\times \U(1)_\rmV}{\mathbb{Z}_{N_f}}, 
\label{eq:symmetry_quark_chemical}
\ee
and the same type of explicit breaking occurs for $\SU(N_f)_{\rmL,\rmR}$ by isospin chemical potential, 
\be
\SU(N_f)_{\rmL,\rmR} \xrightarrow{\,\, \mu_I \,\,} \frac{\SU(N_f/2)\times \SU(N_f/2)\times \U(1)}{\mathbb{Z}_{N_f/2}}.
\label{eq:symmetry_isospin_chemical} 
\ee
Thus, when both $\mu$ and $\mu_I$ are present, we obtain the resulting global symmetry by combining Eqs.~\eqref{eq:symmetry_quark_chemical} and \eqref{eq:symmetry_isospin_chemical}. 

Lastly, let us comment on the sign problem. The $1$-flavor Dirac operator at finite $\mu$, $\mathsf{D}(\mu)=\gamma_\nu D_\nu+\mu\gamma_0$, has the quartet pairing of the spectrum $\{\pm\lambda,\pm\lambda^*\}$ if $\lambda \in \mathbb{C}\setminus \mathbb{R}$ and has the doublet pairing $\{\pm \lambda\}$ if $\lambda \in \mathbb{R}$~\cite{Hands:2000ei}. This is a consequence of the pseudo-reality of $\SU(2)$ and the existence of chiral symmetry. 
Therefore, with the quark mass $m$, the $1$-flavor Dirac determinant takes the form, 
\be
\det(\mathsf{D}(\mu)+m)=\prod_{\lambda\in\mathbb{R}} (m^2-\lambda^2) \prod_{\lambda\not\in\mathbb{R}} |m^2-\lambda^2|^2. 
\ee
This shows that the $1$-flavor Dirac determinant at finite density is real-valued, but does not have to be positive semi-definite. 
Therefore, we need even number of flavors in order to evade the sign problem in Monte Carlo simulation. 

When we consider both the quark and isospin chemical potentials, this sign problem comes back. 
The $u$ quark sector acquires the chemical potential $\mu+\mu_I$, while the $d$ quark sector has $\mu-\mu_I$, so the Dirac determinant, 
\be
\det(\mathsf{D}(\mu+\mu_I)+m)\det(\mathsf{D}(\mu-\mu_I)+m),
\ee
again can oscillate between positive and negative values. Thus, when we introduce both the quark and isospin chemical potentials, the Dirac flavors must be multiples of $4$ to be sign-problem free. 

In order to avoid this problem within two flavors, we can consider real quark chemical potential and imaginary isospin chemical potential, i.e., we replace $\mu_I\to \im \mu'_I \in \im \, \mathbb{R}$. 
Then, using the pseudo-reality of $\SU(2)$, we can show that 
\be
\det(\mathsf{D}(\mu-\im \mu'_I)+m)=\det(\mathsf{D}(\mu+\im \mu'_I)+m)^*. 
\ee
Therefore, the Dirac determinant for the $d$-quark sector is given by the complex conjugate with that of the $u$-quark sector. 
The product of the Dirac determinants over $u,d$ sectors is now positive semi-definite, and it is free from the sign problem for any even $N_f$.

\subsection{Roberge-Weiss Periodicity for Imaginary Isospin Chemical Potentials}
\label{sec:RW_periodicity}

The absence of the sign problem for real quark chemical potentials, $\mu$, and imaginary isospin chemical potentials, $\mu_I=\im \mu'_I$, motivates us to study this setup in more details. 
Especially, we find an isospin version of the Roberge-Weiss (RW) periodicity~\cite{Roberge:1986mm} and an interesting symmetry enhancement for a specific value of the imaginary isospin chemical potential. 

Let $L$ be the length of $S^1$ along the $x^0$ direction\footnote{Throughout this paper, we call $T=1/L$ as the temperature whether the boundary condition is the thermal one or not. }, and then it is convenient to introduce the dimensionless imaginary isospin chemical potential, $\theta_I$, by 
\be
 \mu_I = \im \frac{\theta_I}{ L}. 
\label{eq:dimensionless_isospin}
\ee 
In order to see the RW periodicity, let us redefine the Weyl fermion $\Psi$ as 
\be
\Psi' = \exp\left(-\im \theta_I T_I {x^0\over L}\right)\Psi,\quad 
\overline{\Psi}' =\overline{\Psi}\exp\left(\im \theta_I T_I{x^0\over L}\right). 
\label{eq:redefinition}
\ee
With this redefinition, the imaginary isospin chemical potential is removed from the kinetic term as
\be
\overline{\Psi}\overline{\sigma}^0 \left(\p_0-\mu T_Q-\im {\theta_I\over L}T_I\right)\Psi=\overline{\Psi}' \overline{\sigma}^0 \left(\p_0-\mu T_Q\right)\Psi', 
\ee
and its information is encoded into the phase acquired by the fields along $S^1$,\footnote{Following the standard convention of thermal quantum field theories, we here put the $(-1)$ sign for the fermion boundary condition. We note, however, that the periodic and anti-periodic boundary conditions give the same result for the $\SU(2)$ fundamental fermions, because the difference can be absorbed by gauge transformations. } 
\be
\Psi'(x^0+L) = -\rme^{\im \theta_I T_I}\Psi'(x^0). 
\ee
Naively, this expression tells us that $\theta_I$ is a periodic variable with $\theta_I\sim \theta_I+2\pi$. 
However, the center element $-1$ of $\SU(2)$ color gives the extra identification, 
\be
\theta_I\sim \theta_I+\pi,
\label{eq:RW_periodicity}
\ee
and we call this shortened periodicity as the isospin RW periodicity. 

Let us now discuss the global symmetry. For simplicity of notation, we consider $N_f=2$ in the following, and the generalization for larger even $N_f$ is straightforward. 
Under the presence of $\mu$ and generic values of $\theta_I$, the global symmetry $G_{\mu,\theta_I}$ is 
\be
\begin{split}
G_{\mu,\theta_I}&={\U(1)_{\rmL,3} \times \U(1)_{\rmR,3} \times \U(1)_\rmV\over \mathbb{Z}_2\times \mathbb{Z}_2}
\\
&\subset {\SU(2)_\rmL\times \SU(2)_\rmR \times \U(1)_\rmV\over \mathbb{Z}_2\times \mathbb{Z}_2} \subset {\SU(4)\over \mathbb{Z}_2}, 
\end{split}
\label{eq:symmetry_generic}
\ee
where $\U(1)_{\rmL/\rmR,3}$ denote the Cartan subgroups of $\SU(2)_{\rmL/\rmR}$, respectively. 
However, there are two special points of $\theta_I$ mod $\pi$.  
The obvious one is $\theta_I=0$ mod $\pi$, in which case the global symmetry is (\ref{eq:symmetry_quark_chemical}),
\be
G_{\mu,\theta_I=0}={\SU(2)_\rmL\times \SU(2)_\rmR \times \U(1)_\rmV\over \mathbb{Z}_2\times \mathbb{Z}_2}. 
\ee
We note that, for example, not only $\theta_I=0$ but also $\theta_I=\pi$ have this global symmetry as they are gauge equivalent. 

Another special point is 
\be
\theta_I={\pi\over 2},
\ee 
which we would refer as the isospin RW point. 
In this case, let us consider the $\pi$ rotation along $\tau_1\in \SU(2)_\rmV$, which exchanges $u$ and $d$ quarks, 
\be
u \leftrightarrow d, \; \tilde{u}\leftrightarrow \tilde{d}. 
\ee
This effectively flips the sign of the isospin chemical potential, $\theta_I\mapsto -\theta_I$. Thus, at generic values of $\theta_I$, this is not a symmetry. 
However, at $\theta_I=\pi/2$, this is invariant thanks to the isospin RW periodicity, 
\be
\theta_I=\pi/2 \to -\pi/2 \sim \pi/2. 
\ee
We note that the isospin RW periodicity uses the center transformation along the $x^0$ direction. Therefore, this transformation yields a sign factor to the Polyakov loop, 
\be
\tr(P)=\tr\left[\mathcal{P}\exp\left(\im \oint_{S^1} a_0 \diff x^0\right)\right]. 
\ee
Therefore, at the isospin RW point, we have found the $\mathbb{Z}_2$ symmetry, which acts both on quark flavors and the Polyakov loop at the same time as follows:
\be
\Psi=\begin{pmatrix}
u\\
d\\
\tilde{u}\\
\tilde{d}
\end{pmatrix} \mapsto 
\begin{pmatrix}
d\\
u\\
\tilde{d}\\
\tilde{u}
\end{pmatrix}, \quad \tr(P)\mapsto -\tr(P). 
\ee
This is the same with the $\mathbb{Z}_N$ center symmetry identified in $\mathbb{Z}_N$-twisted QCD for $N_c=N_f=N$~\cite{Kouno:2012zz, Sakai:2012ika, Kouno:2013zr, Kouno:2013mma, Poppitz:2013zqa, Iritani:2015ara, Kouno:2015sja, Hirakida:2016rqd, Hirakida:2017bye, Cherman:2017tey}\footnote{We note that the same/similar boundary conditions play an important role to study the ground-state structures of strongly-coupled theories via adiabatic continuity to weakly-coupled regions~\cite{Tanizaki:2017qhf, Smilga:1993sn, Shifman:1994ce,  Dunne:2012ae,Unsal:2007jx,Unsal:2008ch, Unsal:2007vu, Kovtun:2007py,  Shifman:2008ja, Shifman:2009tp, Cossu:2009sq, Cossu:2013ora, Argyres:2012ka, Argyres:2012vv, Dunne:2012zk, Poppitz:2012sw, Anber:2013doa, Basar:2013sza, Cherman:2014ofa, Misumi:2014raa, Misumi:2014jua, Misumi:2014bsa, Dunne:2015ywa,Misumi:2016fno, Cherman:2016hcd, Fujimori:2016ljw, Fujimori:2017oab, Fujimori:2017osz, Fujimori:2018kqp, Sulejmanpasic:2016llc, Yamazaki:2017ulc, Itou:2018wkm, Buividovich:2017jea, Aitken:2017ayq, Fujimori:2019skd}. }, so we here call it as $(\mathbb{Z}_2)_{\mathrm{center}}$. 
Since this $(\mathbb{Z}_{2})_\mathrm{center}$ flips the charges of $\U(1)_{\rmL,3}\times \U(1)_{\rmR,3}$, the global symmetry at the isospin RW point takes the form of semi-direct product:
\be
G_{\mu,\theta_I=\pi/2}={(\mathbb{Z}_2)_{\mathrm{center}} \ltimes [\U(1)_{\rmL,3} \times \U(1)_{\rmR,3}] \times \U(1)_\rmV\over \mathbb{Z}_2\times \mathbb{Z}_2}, 
\label{eq:RW_symmetry}
\ee
i.e., $G_{\mu,\theta_I=\pi/2}=(\mathbb{Z}_2)_{\mathrm{center}}\ltimes G_{\mu,\theta_I\not =0,\pi/2}$. 
Because the $(\mathbb{Z}_2)_{\mathrm{center}}$ symmetry at $\theta_I=\pi/2$ acts on the Polyakov loop, it deserves the special attention, and we shall discuss properties at the isospin RW point in more detail in later sections.

\subsection{Chiral Effective Lagrangian in $2$-Flavor Case}\label{sec:chiral_EFT_Nf2}

For $N_f=2$, we try to see how the global symmetry is realized in the low-energy chiral Lagrangian~\cite{Kogut:1999iv, Kogut:2000ek}. 
We note that there is an exceptional isomorphism, $\SU(4)\simeq \mathrm{Spin}(6)$ and $\Sp(2)\simeq \mathrm{Spin}(5)$ (see Appendix~\ref{sec:symplectic}), so the chiral symmetry breaking (\ref{eq:chiral_SSB_vacuum_general}) is realized as 
\be
\SO(6)\to \SO(5). 
\label{eq:chiral_SSB_vacuum}
\ee
The order parameter field, $\Sigma=\Psi\Psi$, can be mapped to the $6$-dimensional unit vectors $\bm{n}\in S^5\subset \mathbb{R}^6$, i.e., $\bm{n}\cdot \bm{n}=1$. 
The physical interpretation of this unit vector is as follows: 
\be
\bm{n}=\begin{pmatrix}
n_1\\ n_2\\ n_3\\ n_4\\ n_5\\ n_6
\end{pmatrix}
=\begin{pmatrix}
\sigma\\
\pi_0\\
\pi_1\\
\pi_2\\
\Delta_1\\
\Delta_2
\end{pmatrix}
=\begin{pmatrix}
(\tilde{u}u+\tilde{d}d)/\sqrt{2}\\
(\tilde{u}u-\tilde{d}d)/\sqrt{2}\\
(\tilde{u}d+\tilde{d}u)/2\\
(\tilde{u}d-\tilde{d}u)/2\im \\
(ud+\tilde{d}\tilde{u})/2\\
(ud-\tilde{d}\tilde{u})/2\im
\end{pmatrix}. 
\label{eq:definition_sigma_model}
\ee
That is, $\sigma$ is the sigma meson, which is an isospin singlet, $\vec{\pi}=(\pi_0,\pi_1,\pi_2)$ are pions, which form an isospin triplet, and $\vec{\Delta}=(\Delta_1,\Delta_2)$ are diquarks, which are isospin singlets with baryon-charge $1$. 
The leading term of the chiral Lagrangian is given as 
\be
{f_\pi^2\over 2}\p_\nu \bm{n}\cdot \p_\nu \bm{n}={f_\pi^2\over 2}(|\p_\nu\sigma|^2+|\p_\nu\vec{\pi}|^2+|\p_\nu \vec{\Delta}|^2), 
\ee
where $f_\pi$ denotes the pion decay constant. 

When we add a Dirac mass, $m>0$, it gives the explicit breaking term, 
\be
- f_\pi^2m \Lambda \sigma, 
\ee
where $\Lambda$ is a strong scale ($f_\pi$ and $\Lambda$ are of the same order). In this case, the vacuum is chosen to be
\be
\langle \sigma\rangle=1,
\ee
and other fields are zero. Pions and diquarks acquire the same mass $m_\pi=\sqrt{m\Lambda}$, which obeys the Gell-Mann--Oakes--Renner relation. 

Next, we introduce the quark chemical potential. The quark number rotation acts as $(u,d)\to \rme^{\im \alpha}(u,d)$ and $(\tilde{u},\tilde{d}) \to \rme^{-\im\alpha}(\tilde{u},\tilde{d})$, 
so $\bm{n}$ transforms as 
\be
\sigma\to \sigma,\; \vec{\pi} \to \vec{\pi},\; \vec{\Delta}\to \exp\left(2\im\alpha \tau_2\right)\vec{\Delta}. 
\ee
Under the presence of quark chemical potential, we replace $\p_0\vec{\Delta}$ by $(\p_0+2\mu\tau_2)\vec{\Delta}$. So, the term $|\p_0\vec{\Delta}|^2$ now becomes 
\be
\begin{split}
&([\p_0+2\mu\tau_2]\vec{\Delta})^T([\p_0+2\mu\tau_2]\vec{\Delta})
\\
&=|\p_0\vec{\Delta}|^2+2\mu (\p_0\vec{\Delta}\cdot \tau_2\vec{\Delta}-\vec{\Delta}\cdot \tau_2\p_0\vec{\Delta})-(2\mu)^2\vec{\Delta}^2. 
\end{split}
\ee
Therefore, at nonzero quark chemical potential, the vacuum manifold is $S^1$ given by 
\be
\vec{\Delta}^2=1. 
\label{eq:diquark_condensate}
\ee
The pions and sigma meson acquire the mass $2|\mu|$~\cite{Kogut:1999iv, Kogut:2000ek}. 
The diquark field $\vec{\Delta}$ is singlet under isospin chiral symmetry, $\SU(2)_\rmL\times \SU(2)_\rmR$, and the pattern of symmetry breaking is 
\be
{\SU(2)_\rmL\times \SU(2)_\rmR\times \U(1)_\rmV\over \mathbb{Z}_2\times \mathbb{Z}_2}\to {\SU(2)_\rmL\times \SU(2)_\rmR\over \mathbb{Z}_2}. 
\ee
The isospin chemical potential does the same job. Instead of replacing $\p_0 \vec{\Delta}$, we have to replace 
\be
\p_0 \begin{pmatrix}\pi_1\\ \pi_2\end{pmatrix} \Rightarrow [\p_0+2\mu_I \tau_2]\begin{pmatrix}\pi_1\\ \pi_2\end{pmatrix}.
\ee 
Especially, we are interested in the case with the imaginary isospin chemical potential $\mu_I=\im \theta_I/L$. This increases the energy for $(\pi_1,\pi_2)$ direction unlike the case of the real chemical potential. 

When we introduce the quark chemical potential and imaginary isospin chemical potential under the presence of the small quark mass, we obtain the effective potential\footnote{For simplicity of the expression, we here pick up only Matsubara zero modes of mesons and diquarks. We note that, in this truncation, the isospin RW periodicity (\ref{eq:RW_periodicity}) is violated. This can be fixed by reinstating all the Matsubara frequencies for $\pi_{1,2}$. As a consequence, charged pions $\pi^{\pm}=(\pi_1\pm\im \pi_2)$ of Matsubara frequency $\omega_n={2\pi\over L}n$ get the mass, $|\omega_n\pm 2\theta_I/L|$, and $\theta_I\to \theta_I+\pi$ gives the level crossing of Matsubara modes, respecting the isospin RW periodicity. As they acquire positive energies, the consequence is unaffected by these details. 
}, 
\be
\frac{V_\mathrm{eff}}{f_\pi^2 m_\pi^2}= -\sigma - \frac{1}{2}\left({2\mu\over m_\pi}\right)^2 \vec{\Delta}^2
+\frac{1}{2} \left({2\theta_I\over L m_\pi}\right)^2 (\pi_1^2+\pi_2^2). 
\label{eq:effective_potential}
\ee
Setting $\vec{\pi}=0$, we can analyze the minima of this potential by substituting $\sigma=\sqrt{1-\vec{\Delta}^2}$. 
We can find the second-order phase transition at $\mu=\mu_c=m_\pi/2$~\cite{Kogut:1999iv, Kogut:2000ek},  where $\langle \sigma\rangle=1$ for $\mu<\mu_c$. 
So, diquark condensation starts to appear at the half of pion mass. 
In the chiral limit, this critical value goes to zero, and the diquark condensate and chiral condensate describe the same symmetry breaking pattern at $\mu=0$. 

\section{Perturbative 't~Hooft Anomaly of Massless QC$_2$D}\label{sec:perturbative_anomaly}

In this and next sections, we study the nature of 't~Hooft anomalies in massless QC$_2$D. 
An 't~Hooft anomaly can be characterized as an obstruction of gauging global symmetries, and, importantly, this anomaly is invariant under the renormalization group (RG) flow~\cite{tHooft:1979rat, Frishman:1980dq, Coleman:1982yg}. 
Because of this RG invariance, 't~Hooft anomaly provides a useful constraint on low-energy dynamics of strongly-coupled systems, and this is called the anomaly matching condition. 
In this section, we compute the 't~Hooft anomaly for infinitesimal chiral transformations, i.e., a perturbative anomaly. 
This anomaly is already useful to constrain the properties of QC$_2$D dynamics at the zero temperature ($T=0$ or $L=\infty$). 
In the next section, we shall discuss a more subtle 't~Hooft anomaly. 

\subsection{Perturbative Anomaly of $\SU(2N_f)$ for $\mu=0$}

Let us compute the perturbative anomaly of $\SU(2N_f)$ chiral symmetry. Since the discrete factor does not affect the computation of the perturbative anomaly, we will be ignorant about it in this section. Its subtle effect is taken into account in the next section and essential in the computation of discrete anomaly. 

To see the existence of anomaly, we introduce the $\SU(2N_f)$ background gauge field $A$, and replace the quark kinetic term as 
\be
\overline{\Psi} \overline{\sigma}^\nu D_\nu \Psi \; \Rightarrow \; \overline{\Psi} \overline{\sigma}^\nu (D_\nu+\im A_\nu) \Psi.
\ee
In this way, we can compute the partition function $Z[A]$ with the $\SU(2N_f)$ background gauge field $A$, but this partition function does not have to be gauge invariant for $A$. 
The anomaly can be explicitly computed by the Fujikawa method~\cite{Fujikawa:1979ay, Fujikawa:1980eg, AlvarezGaume:1984dr}, but it can also be obtained in ad hoc way by the Stora--Zumino descent procedure~\cite{Stora:1983ct, Zumino:1983ew} since the anomaly must satisfy the Wess--Zumino (WZ) consistency condition~\cite{Wess:1971yu}. 
Here, we choose to use the anomaly descent procedure.
Let us start from the $6$-dimensional Abelian anomaly:
\be
2\times {2\pi\over 3! (2\pi)^3}\tr (F_A^3), 
\label{eq:6d_anomaly_PG}
\ee
where $F_A=\diff A+\im A\wedge A$, and the factor $2$ in front comes from the number of color. 
This gives the $5$-dimensional parity anomaly, characterized by the Chern--Simons action, 
\be
2 \, \mathrm{CS}_5[A], 
\ee
where $\diff (\mathrm{CS}_5[A])={1\over 24\pi^2}\tr (F_A^3)$. This topological action completely specifies the perturbative anomaly of $Z[A]$; the system $Z[A]$ as the boundary of $5$d Chern--Simons theory,
\be
Z[A]\exp\left(2\im\, \int \mathrm{CS}_5[A]\right), 
\ee
is gauge invariant, because the boundary term for gauge variations of $2\,\mathrm{CS}_5[A]$ cancels the anomaly of $Z[A]$. 

When $N_f$ is not too large, it is natural to expect that anomaly matching is satisfied by chiral symmetry breaking, $\SU(2N_f)\to H$. 
If we further assume that anomaly matching is satisfied only by NG bosons, $H$ must be anomaly free. 
There are two important anomaly-free subgroups of $\SU(2N_f)$, which are $H=\SO(2N_f)$\footnote{Spontaneous breaking, $\SU(2N_f)\to \SO(2N_f)$, occurs if the gauge group is strictly real instead of pseudo-real, because the chiral condensate, $\Psi\Psi$, is then in the two-index symmetric representation~\cite{Peskin:1980gc}. Because of the exceptional isomorphism, $\mathrm{Lie}(\SO(6))=\mathrm{Lie}(\SU(4))$, however, we cannot immediately say $\SO(6)\subset \SU(6)$ is anomaly free, so one may wonder if $N_f=3$ can be special. But, this does not happen fortunately. For $N_f=3$, $\Psi$ is in $\bm{6}$ representation of $\SU(6)$, which is in the two-index anti-symmetric representation of $\SU(4)(=\mathrm{Spin}(6))$ and does not have the triangle anomaly.  } and $H=\Sp(N_f)$. 
Indeed, for QC$_2$D, we are expecting the spontaneous breaking pattern is $\SU(2N_f)\to \Sp(N_f)$, as we have reviewed in Sec.~\ref{sec:Pauli_Gursey}. 
We can match the anomaly with this symmetry breaking pattern using the WZ term, and its explicit form can be found, e.g., in Ref.~\cite{Brauner:2018zwr}. 

\subsection{Perturbative Anomaly of $\SU(N_f)_\rmL\times \SU(N_f)_\rmR\times \U(1)_\rmV$ for $\mu>0$}

We move on to the discussion of the perturbative anomaly for $\mu>0$. The perturbative anomaly matching at $T=0$ and finite $\mu$ has been considered in Ref.~\cite{Sannino:2000kg} for $N_c=3$, and we discuss it here in the context of QC$_2$D. 

As we have seen in (\ref{eq:symmetry_quark_chemical}), the $\SU(2N_f)$ chiral symmetry is explicitly broken to $\SU(N_f)_\rmL\times \SU(N_f)_\rmR \times \U(1)_\rmV$ by the presence of $\mu$. 
Let us denote $(F_\rmL,F_\rmR,F_\rmV)$ as the gauge-field strengths of $\SU(N_f)_\rmL\times \SU(N_f)_\rmR \times \U(1)_\rmV$, then they are embedded into the $\SU(2N_f)$ field strength, $F_A$, as follows: 
\be
F_A=\begin{pmatrix}
F_\rmL+F_\rmV & \bm{0}\\
\bm{0} & -(F_\rmR+F_\rmV)
\end{pmatrix},
\ee 
according to (\ref{eq:embedding_chiral_symmetry}). 
Substituting this expression into (\ref{eq:6d_anomaly_PG}), we can obtain the $6$-dimensional form for the Stora--Zumino procedure, 
\be
\begin{split}
{2\over 24\pi^2}\tr(F_A^3)=&{2\over 24\pi^2}\tr (F_\rmL^3-F_\rmR^3) 
\\
&+ {2\over 8\pi^2} F_\rmV\wedge \tr(F_\rmL^2-F_\rmR^2). 
\end{split}
\label{eq:6d_anomaly_LR}
\ee
Let us now discuss how we can match the anomaly from spontaneous symmetry breaking. 

A typical example of anomaly-free subgroups is the vector-like subgroup, which leads to the standard chiral symmetry breaking of QCD with $N_c\ge 3$, $\SU(N_f)_\rmL\times \SU(N_f)_\rmR\to \SU(N_f)_\rmV$. Indeed, once we assume this spontaneous breaking pattern, we can match both terms of the anomaly (\ref{eq:6d_anomaly_LR}) at once. 

However, QC$_2$D has the massless diquarks, which start to condense immediately after introducing nonzero $\mu$. 
This condensation breaks $\U(1)_\rmV$ spontaneously, so the second term of (\ref{eq:6d_anomaly_LR}) is already matched by the associated NG boson. 
Therefore, $\SU(N_f)_\rmL\times \SU(N_f)_\rmR$ has to be broken so as to match only the first term of (\ref{eq:6d_anomaly_LR}). 
This requirement opens a new possibility to saturate the anomaly. 
Indeed, when $N_f$ is even, one of the possible patterns is 
\be
\begin{split}
\SU(N_f)_\rmL\times& \SU(N_f)_\rmR \times \U(1)_\rmV\,
\\
&\to\, \Sp(N_f/2)_\rmL\times \Sp(N_f/2)_\rmR.  
\end{split}
\ee
This breaking pattern is indeed found in the analysis of chiral effective model~\cite{Kogut:1999iv, Kogut:2000ek}. The number of NG bosons is $N_f^2-N_f-1$. 

Especially when $N_f=2$, since $\SU(2)=\Sp(1)$, it shows no chiral symmetry breaking at finite densities, and we only have $\U(1)_\rmV/\mathbb{Z}_2 \to 1$, as we have discussed in Sec.~\ref{sec:chiral_EFT_Nf2}. 
We note that this is consistent with (\ref{eq:6d_anomaly_LR}), because $\tr(F_\rmL^3)=\tr(F_\rmR^3)=0$ for $N_f=2$. 
Within the chiral effective description, the vacuum manifold (\ref{eq:diquark_condensate}) can be parametrized by the $2\pi$ periodic scalar field, $\varphi$, where 
\be
\Delta_1+\im \Delta_2=\rme^{\im \varphi}. 
\ee
Under the presence of background gauge fields, we can write down the axion-like coupling,
\be
{\im \over 8\pi^2} \varphi \wedge \tr(F_\rmL^2-F_\rmR^2). 
\ee
It is exactly the term that matches the second term of the anomaly (\ref{eq:6d_anomaly_LR}), since $\varphi$ has the quark charge $2$, i.e., the baryon charge $1$. 


\section{Discrete Anomaly of Massless $2$-Flavor QC$_2$D at Isospin RW Point}\label{sec:discrete_anomaly}

In this section, we discuss a more subtle anomaly related to the discrete factor of the global symmetry. 
The discussion here is a crucial step to find a rigorous constraint on the phase diagram of QC$_2$D at the isospin RW point. 
Perturbative anomalies, discussed in the previous section~\ref{sec:perturbative_anomaly}, is very powerful for restrictions on the massless spectrum at $T=0$, but they do not provide useful constraints  on the phase diagram. 
This can be seen from the well-known fact that the high-temperature phase of QCD with fundamental quarks is a trivial phase: The vacuum is unique, quark excitations are gapped by Matsubara frequencies ($\ge \pi T$), and gluon excitations are also gapped because of the electric ($\sim gT$) and magnetic ($\sim g^2 T$) masses. 

When we introduce a specific flavor-twisted boundary condition, however, it is no longer the case. The high-temperature phase is also nontrivial as it is doubly degenerate by the RW phase transition~\cite{Roberge:1986mm}, for example. 
Similar vacuum degeneracy is now found in other systems and understood as consequences of anomaly matching of subtle discrete anomalies~\cite{Gaiotto:2017yup, Shimizu:2017asf, Tanizaki:2017qhf, Tanizaki:2017mtm, Dunne:2018hog, Yonekura:2019vyz}rather than perturbative anomalies. 
In this section, we, therefore, compute the discrete anomaly for massless two-flavor QC$_2$D at the isospin RW point.

In order to see the discrete anomaly in our system, we have to introduce the background gauge field for $G_{\mu,\theta_I=\pi/2}$ in (\ref{eq:RW_symmetry}) very carefully, as we can easily miss such an anomaly just by getting an extra factor $2$. 
For this purpose, it is convenient to rewrite the global symmetry to eliminate the redundancy as much as possible. 
Indeed, the symmetry group at generic values of $\mu$ and $\theta_I$, $G_{\mu,\theta_I}$ in (\ref{eq:symmetry_generic}), can be written as 
\begin{align}
G_{\mu,\theta_I}=\frac{ \U(1)^{\rmu}_{\rmV} \times \U(1)^\rmd_{\rmV}}{\Z_2}\times \U(1)_{\rmL,3}.
\label{eq:symmetry_generic_simplify}
\end{align}
Here, $\U(1)^{\rmu,\rmd}_\rmV$ are the vector-like $\U(1)$ symmetries acting on $u$ and $d$ quarks, respectively. 
They act on the quark fields as
\be
\begin{pmatrix}
u \\ d \\ \tilde{u} \\ \tilde{d}
\end{pmatrix}
		\ra 
		\begin{pmatrix}
		 \rme^{\im \alpha_{\rml,3}+\im \alpha^\rmu_{\rmv}} & \  & \ & \
		 \\
		 \ & \rme^{-\im \alpha_{\rml,3}+\im \alpha^\rmd_{\rmv}}   & \ & \
		 \\
		 \ & \  & \rme^{ - \im \alpha^\rmu_{\rmv}} & \
		 \\
		 \ & \  & \ & \rme^{-\im \alpha^\rmd_{\rmv}}
		\end{pmatrix}
		\begin{pmatrix} 
		u \\ d \\ \tilde{u} \\ \tilde{d}
		\end{pmatrix},
\ee
where $\rme^{\im \alpha_{\rml,3}}$, $\rme^{\im \alpha^\rmu_{\rmv}}$, and $\rme^{\im \alpha^\rmd_{\rmv}}$ belong to $\U(1)_{\rml,3}$, $\U(1)^\rmu_{\rmv}$, and $\U(1)^\rmd_{\rmv}$, respectively.
These three $\U(1)$ transformations do not have any overlaps, which is why we have only one $\mathbb{Z}_2$ quotient in (\ref{eq:symmetry_generic_simplify}) rather than two\footnote{Recall that the previous expression (\ref{eq:symmetry_generic}) has two $\mathbb{Z}_2$ quotients because of an extra overlap between quark-number and isospin symmetries.}.

As discussed around (\ref{eq:RW_symmetry}), the model acquires the $(\mathbb{Z}_2)_{\mathrm{center}}$ symmetry at the isospin RW point. This symmetry flips the sign of the $\U(1)_{\rmL,3}$ charge while it exchanges the charges of $\U(1)^{\rmu}_{\rmv}$ and $\U(1)^{\rmd}_{\rmv}$.
We try to find the $\mathbb{Z}_2$ anomaly by two steps: We first introduce the background gauge field for $G_{\mu,\theta_I}$ in a gauge-invariant way, and then we shall observe the violation of $(\mathbb{Z}_2)_{\mathrm{center}}$~\cite{Yonekura:2019vyz}. 
Our discussion is analogous to the parity anomaly of three-dimensional Dirac fermions.

Introducing the background gauge field for $G_{\mu,\theta_I}$, the gauge structure becomes 
\be
{\SU(2)_{\mathrm{gauge}}\times \U(1)^\rmu_\rmV\times \U(1)^\rmd_\rmV\over \mathbb{Z}_2}\times \U(1)_{\rmL,3}. 
\ee
We denote the background gauge fields for the $\U(1)_{\rmL,3}$, $\U(1)^{\rmu}_{\rmv}$, and $\U(1)^\rmd_{\rmv}$ symmetries as $A_{\rml,3}$, $A^{\rmu}_{\rmv}$, and $A^{\rmd}_{\rmv}$. 
Because of the $\mathbb{Z}_2$ quotient, we also need to introduce a $\mathbb{Z}_2$ two-form gauge field $B$ and must postulate the invariance under one-form gauge transformation, $B \to B+\diff \Lambda$~\cite{Kapustin:2014gua}. 
Note that in this basis, the $\Z_2$ one-form gauge invariance is implemented as
\begin{align}
A^{\rmu}_{\rmv} &\ra A^{\rmu}_{\rmv} - \Lambda,
\\
A^{\rmd}_{\rmv} &\ra A^{\rmd}_{\rmv} - \Lambda.
\end{align}
Therefore, one-form gauge invariant combinations of their field strengths are given by 
\be
F_\rmu=\diff A^\rmu_\rmV+B,\quad F_\rmd=\diff A^\rmd_\rmV+B, 
\label{eq:one_form_gaugeinv_UD}
\ee
and we can identify the gauge field $A_\rmB$ for the baryon-number (not quark-number) symmetry as~\cite{Tanizaki:2018wtg}
\be
\diff A_\rmB=F_\rmu+F_\rmd=\diff A^\rmu_\rmV+\diff A^\rmd_\rmV+2B. 
\label{eq:baryon_number_gauge}
\ee
This baryon-number gauge field, $A_\rmB$, satisfies the canonical geometric normalization of the $\U(1)$ gauge field. 
For later use, we mention that the $(\Z_2)_{\mathrm{center}}$ symmetry acts on the background gauge fields as
\be
A_{\rml,3} \ra -A_{\rml,3},\; \;
A^{\rmu}_{\rmv} \longleftrightarrow A^{\rmd}_{\rmv},\; \;
B \ra B.
\label{eq:z2centeraction}
\ee
Especially, we note that $A_\rmB$ is unchanged under $(\mathbb{Z}_2)_{\mathrm{center}}$.

Let all the background gauge field be three-dimensional ones so that we discuss an anomaly present even in the high-temperature limit.
To understand how the background gauge fields violate the $(\mathbb{Z}_2)_{\mathrm{center}}$ symmetry,
recall the derivation on the $(\Z_2)_\mathrm{center}$ symmetry in Sec~\ref{sec:RW_periodicity}.
The key step in the discussion is translating the isospin imaginary chemical potential into the twisted boundary condition [see Eq.~\eqref{eq:redefinition}].
When the background gauge fields are turned on, however,
the fermion path integral measure generates the following phase factor under this operation:
\be\label{eq: s'}
S' ={\im \over 2} \int_3 \frac{1}{2 \pi} A_{\rml,3}\wedge \diff A_\rmB,
\ee
which one can readily show via the standard Fujikawa method for chiral gauge theories~\cite{Fujikawa:2004cx}.
Thus, we must take into account the extra phase factor $S'$ when we consider the $(\Z_2)_\mathrm{center}$ symmetry in the presence of the background gauge fields.

The generated phase factor $S'$ is actually responsible for the 't Hooft anomaly of QC$_2$D at the isospin RW point.
Let us perform the $(\Z_2)_{\mathrm{center}}$ transformation~\eqref{eq:z2centeraction}.
This keeps the original action unchanged, but due to the generated phase factor, it changes the partition function as follows:
\begin{align}
Z[A_{\rmL,3} ,A_\rmB] \ra Z[A_{\rmL,3} ,A_\rmB] \rme^{\rmi \mathcal{A}[A_{\rmL,3},A_\rmB]}.
\end{align}
Here, $\mathcal{A}[A_{\rmL,3} ,A_\rmB]$ takes the form:
\be
\label{eq: anom}
\mathcal{A}[A_{\rmL,3},A_\rmB] =
- \frac{1}{2 \pi} \int A_{\rml,3} \wedge  \diff A_\rmb.
\ee
We have no local counter term to cancel this anomaly without breaking the $G_{\mu,\theta_I}$ gauge invariance.
To recover the $(\Z_2)_{\mathrm{center}}$ symmetry keeping the $G_{\mu,\theta_I}$ gauge invariance,
we have to attach QC$_2$D onto the $4$-dimensional bulk mixed theta term, whose theta angle $\pi$:
\be
S_\Theta = \rmi \pi \int_4 \frac{\diff A_{\rmL,3}}{2 \pi}\wedge \frac{\diff A_\rmB}{2 \pi}. 
\label{eq:thetaterm}
\ee
Therefore, we find the anomaly for $G_{\mu,\theta_I=\pi/2}$,
and its classification is $\Z_2$ because it is saturated by the $4$-dimensional mixed theta term.
We would like to emphasize that the anomaly is present on the entire $(\mu, T)$ phase diagram at $\theta_I=\pi/2$.

Let us comment on another understanding of the anomaly.
When we compactify the imaginary time direction, we find an infinite tower of three-dimensional Dirac fermions, whose real masses are given by background Polyakov-loop phases and Matsubara frequencies. 
It is known that, in this setup, we can make the system gauge invariant, e.g., by taking the Pauli--Villars (PV) regularization.
However, such PV regulators induce Chern--Simons terms~\cite{Niemi:1983rq, Redlich:1983dv}, and this is crucial for the parity anomaly. 
In our case, we introduce the isospin chemical potential, $\theta_I=\pi/2$, which is nothing but the zeroth component of the background gauge fields, 
\be
A^\rmu_{\rmV, \nu=0}={\pi\over 2}{1\over L},\;\;  A^\rmd_{\rmV, \nu=0}=-{\pi\over 2}{1\over L}. 
\ee
Because of the shift of real masses between $u$- and $d$-quark sectors, Chern-Simons term is induced by the loop effect, or by the fermion measure, and it ends up with 
\begin{align}
S' ={\im \over 2} \int_3 \frac{1}{2 \pi} A_{\rml,3}\wedge \diff A_\rmB.
\end{align}
It term coincides with Eq.~\eqref{eq: s'}.

Moreover, we can make the connection of this loop-induced Chern-Simons term and the perturbative anomaly discussed in Sec.~\ref{sec:perturbative_anomaly} (see also Refs.~\cite{Yonekura:2019vyz, Poppitz:2008hr}). 
Our contents of gauge fields can be embedded into the $\SU(4)$ chiral gauge field in Sec.~\ref{sec:perturbative_anomaly} as 
\be
F_A=\mathrm{diag}\left(F_\rmu+\diff A_{\rmL,3}, F_\rmd-\diff A_{\rmL,3}, -F_\rmu, -F_\rmd\right),
\ee
where $F_{\rmu,\rmd}$ are given by (\ref{eq:one_form_gaugeinv_UD}) and the one-form gauge invariance is already taken into account. The $6$-dimensional form of the Stora-Zumino descent procedure now becomes 
\be
\begin{split}
2\, {2\pi\over 3!(2\pi)^3}\tr(F_A^3)={1\over (2\pi)^2} \diff A_{\rmL,3} \wedge \diff A_\rmB \wedge (F_\rmu-F_\rmd)&
\\
+{1\over (2\pi)^2} (\diff A_{\rmL,3})^2 \wedge \diff A_\rmB&.  
\end{split}
\label{eq:SZ_discrete_6d}
\ee
The significant term for us is the first one on the right hand side. The effect of the imaginary isospin chemical potential can be expressed as the imaginary-time integration along $S^1\ni x^0$ of the background gauge fields:
\be
\int_{D^2} (F_\rmu-F_\rmd)=\int_{S^1} (A^{\rmu}_V-A^{\rmd}_\rmV)={\pi\over 2}-\left(-{\pi\over 2}\right)=\pi,
\ee
where $D^2$ is the two-dimensional disk whose boundary is the imaginary-time circle $S^1$.
Therefore, this replacement on third component in the first term of (\ref{eq:SZ_discrete_6d}) gives 
\be
{1\over 2}\times {1\over2\pi} \diff A_{\rmL,3}\wedge \diff A_\rmB,
\ee
which is equivalent to the bulk theta term~\eqref{eq:thetaterm}.
This is the characterization of the induced Chern-Simons term~\eqref{eq: s'} via the descent procedure.

\section{Phase Diagram of Massless $2$-Flavor QC$_2$D at Isospin RW Point}
\label{sec:phase_diagram}

We can now discuss the constraint on the phase diagram of QC$_2$D by using the anomaly matching condition. 
At $T=0$, perturbative anomalies in Sec.~\ref{sec:perturbative_anomaly} require the existence of massless excitations. 
At $T\not=0$, those perturbative constraints no longer exist. At the isospin RW point, $\theta_I=\pi/2$, we have the discrete anomaly discussed in Sec.~\ref{sec:discrete_anomaly} and this discrete anomaly gives the constraint on possible phase diagrams even at nonzero $T$~\cite{Shimizu:2017asf, Tanizaki:2017qhf, Tanizaki:2017mtm}. 
In this section, we discuss the possible phase structure consistent with these anomalies, taking into account the analytic results of the chiral effective theory and the numerical results of lattice simulations.

The 't~Hooft anomaly matching argument~\cite{tHooft:1979rat, Frishman:1980dq, Coleman:1982yg} states that the 't Hooft anomaly is preserved under the renormalization group flow so that its low-wavelength effective field theory must reproduce the same 't Hooft anomaly.
As a corollary of the anomaly matching, QC$_2$D must show
\textit{spontaneous symmetry breaking},
\textit{topological order}, or
\textit{conformal behavior},
and a scenario with a unique gapped vacuum is ruled out.
Since the discussion only depends on the symmetry consideration, 
it is applicable massless QC$_2$D with arbitrary temperatures $T=1/L$ and chemical potentials $\mu$, when we introduce the isospin imaginary chemical potential, $\theta_I=\pi/2$.

In order to match the anomaly, we assume that spontaneous symmetry breaking happens to the anomaly-free subgroups. 
That is, at any temperatures and chemical potentials, massless QC$_2$D at isospin RW point at least spontaneously breaks either of 
\begin{itemize}
\item center symmetry [$(\mathbb{Z}_2)_{\mathrm{center}}\to 1$], 
\item chiral symmetry [$\U(1)_{\rmL,3}\to 1$], or
\item baryon-number symmetry [$\U(1)_\rmV/\mathbb{Z}_2\to 1$]. 
\end{itemize}
The anomaly itself allows other possibilities, but the analysis from chiral effective Lagrangian and lattice simulations~\cite{Hands:2006ve, Braguta:2016cpw, Bornyakov:2017txe, Astrakhantsev:2018uzd, Iida:2019rah, Boz:2019enj} supports this assumption. 
The expected phase diagram is given in the right panel of Fig.~\ref{fig:phase-diagram} in the Introduction. We now look into its details and see how the anomaly matching condition is satisfied in each phase.

\subsection{Chiral Symmetry Breaking at $\mu=0$}

Let us first discuss the case $\mu=0$. When the temperature $T$ is not so large, we expect that chiral symmetry breaking is realized. 
We note, however, that chiral symmetry breaking does not have to be caused by the common chiral condensate, $\tilde{\psi}\psi$, at the massless point: The diquark condensate, $\psi\psi$, plays the same role because of the Pauli--G\"{u}rsey symmetry, $\SO(6)$. 

At $T=0$, the isospin chemical potential can be eliminated, so the system is the same with the usual vacuum. In this case, as we have reviewed in Sec.~\ref{sec:symmetry}, chiral symmetry breaking occurs as 
\be
\SO(6)\to \SO(5). 
\ee 
The target space of the nonlinear sigma model is $\bm{n}\in S^5$, which is defined as (\ref{eq:definition_sigma_model}), and the unit vector $\bm{n}$ consists of mesons $(\sigma,\pi_0,\pi_1,\pi_2)$ and diquarks $(\Delta_1,\Delta_2)$. 

Let us now consider nonzero temperatures, $T\not=0$, and then the imaginary isospin chemical potential affects the symmetry. 
Because of the isospin chemical potentials, the $\SO(6)$ chiral symmetry is explicitly broken as 
\be
\SO(6)\xrightarrow{\theta_I} \SO(4). 
\ee
Indeed, within chiral effective Lagrangian (\ref{eq:effective_potential}), the imaginary isospin chemical potential introduces the mass to $\pi_1$ and $\pi_2$: 
\be
V_{\mathrm{eff}}={f_\pi^2 m_\pi^2\over 2}\left({2\theta_I\over L m_\pi}\right)^2 (\pi_1^2+\pi_2^2). 
\ee
Therefore, $\SO(4)$ chiral rotations act on $(\sigma,\pi_0,\Delta_1,\Delta_2)\in S^3\subset \mathbb{R}^4$, as we can set $\pi_1=\pi_2=0$ to minimize the potential. 
Spontaneous symmetry breaking occurs as 
\be
\SO(4)\to \SO(3). 
\ee
We should note that, at $\mu=0$, this symmetry breaking can be regarded both as chiral symmetry breaking, $\U(1)_{\rmL,3}\to 1$, and as baryon-number symmetry breaking, $\U(1)_\rmV/\mathbb{Z}_2\to 1$. Indeed, $\U(1)_{\rmL,3}\times \U(1)_\rmV/\mathbb{Z}_2\subset \SO(4)$, and these two symmetry-breaking patterns are identical up to an $\SO(4)$ chiral transformation.

Let us now ask how the anomaly is matched in this phase. 
We note that this phase hosts a topological soliton because 
\be
\pi_3(\SO(4)/\SO(3))=\pi_3(S^3)=\mathbb{Z}. 
\ee
The topological current takes the same form with that of the Skyrmion current~\cite{Witten:1983tx}, and the effect of the discrete gauge fields has been discussed in \cite{Tanizaki:2018wtg}. An only difference from usual Skyrmions is that the corresponding $\U(1)$ symmetry is the isospin symmetry $\U(1)_{\rmV,3}/\mathbb{Z}_2\subset \SU(2)_\rmV/\mathbb{Z}_2$, not the baryon-number symmetry. 
As it couples to the imaginary isospin chemical potential, $\theta_I$, we can reproduce the correct discrete anomaly (\ref{eq: anom}) (see Ref.~\cite{Yonekura:2019vyz} for details).

\subsection{Baryon Superfluidity at $\mu>0$}

Assuming the chiral symmetry breaking $\SO(6)\to \SO(5)$ in vacuum, there are massless diquarks, $\Delta_1, \Delta_2$, in massless QC$_2$D. 
Therefore, soon after introducing the chemical potential, $\mu\not = 0$, they start to condense as the effective potential (\ref{eq:effective_potential}) takes the form, 
\be
V_{\mathrm{eff}}=-{f_\pi^2 m_\pi^2\over 2}\left({2\mu\over m_\pi}\right)^2 (\Delta_1^2+\Delta_2^2). 
\ee
The ground state at $\mu\not = 0$ breaks the baryon number symmetry spontaneously, 
\be
\U(1)_\rmV/\mathbb{Z}_2\to 1. 
\ee
We parametrize our vacuum manifold as $\Delta_1+\im \Delta_2=\rme^{\im \varphi}$. We note that we encounter the second-order phase transition in the limit $\mu\to 0$ as other NG bosons reduce their mass, which behaves as $2|\mu|$.

In order to match the anomaly (\ref{eq: anom}), 
a vortex of the diquark must carry a nontrivial quantum number of $\U(1)_{\rml,3}$. 
Let us see this explicitly. 
When we turn on the background gauge fields, $A_\rmB$ and $A_{\rmL,3}$, we introduce the topological coupling, 
\begin{align}
S_\mathrm{top}={\im\over 4\pi}\int (\diff \varphi-A_\rmB)\wedge \diff A_{\rmL,3}. 
\end{align}
We note that this is gauge invariant for $A_\rmB$ and $A_{\rmL,3}$, because $(\diff \varphi-A_\rmB)$ is the minimal coupling and $\diff A_{\rmL,3}$ is the field strength. 
In order to see the $\mathbb{Z}_2$ anomaly, we perform the $(\mathbb{Z}_2)_{\mathrm{center}}$ transformation, and then
\be
\begin{split}
S_\mathrm{top}\mapsto& {\im\over 4\pi}\int (\diff \varphi-A_\rmB)\wedge (-\diff A_{\rmL,3}) 
\\ 
=&S_\mathrm{top}+{\im\over 2\pi}\int A_\rmB\wedge \diff A_{\rmL,3}, \quad (\bmod \; 2\pi\im). 
\end{split}\label{eq:redefinition}
\ee
By integration by parts, this extra phase is nothing but the anomaly (\ref{eq: anom}). 
Besides, the mixed anomaly predicts that, 
when we take $\int_{x^1,x^2} \diff A_\rmb = 2 \pi$, this topological coupling induces the one-dimensional anomalous term:
\begin{align}
\frac{i}{2}\int_{x^3}A_{\rml,3}.
\end{align}
In other words, a half vortex of the diquark $\Delta$ supports a one-dimensional theory with the $(\Z_2)_{\mathrm{center}} \ltimes \U(1)_{\rml,3}$ anomaly.

\subsection{Quark Gluon Plasma}

At sufficiently high temperatures, the perturbative potential for the Polyakov loop prefers the center-broken phase by having nonzero expectation value, $\langle P\rangle \not =0$~\cite{Gross:1980br}. 
At the isospin RW point, the high-temperature effective potential is given as~\cite{Cherman:2017tey, Tanizaki:2017mtm} (see also \cite{KorthalsAltes:1999cp})
\begin{widetext}
\be
V_{\mathrm{eff}}(P)=-{2\over \pi^2 L^4}\sum_{n\ge 1}{1\over n^4}\left(|\tr(P^n)|^2-1\right) 
+ {1\over 4\pi^2 L^4}\sum_{n\ge 1}{(-1)^n\over n^4}\left[\rme^{2n\mu L}\tr(P^{2n})+\rme^{-2n\mu L}\tr(P^{-2n})\right]. 
\label{eq:effective potential for polyakov loop}
\ee
\end{widetext}
The first term on the right hand side is the gluon contribution, and the second term comes from the quark contribution with the imaginary isospin chemical potential, $\theta_I=\pi/2$. 
It is important to note that this potential respects the center symmetry, $P\to -P$, as we have reviewed in Sec.~\ref{sec:RW_periodicity}. 
In order to get an insight of this potential, let us pick up the $n=1$ terms in Eq.~\eqref{eq:effective potential for polyakov loop}.
We then see that it is minimized by $P=\pm \bm{1}_2$. Therefore, we get the symmetry breaking,
\be
(\mathbb{Z}_2)_{\mathrm{center}}\to 1. 
\ee
This corresponds to quark-gluon plasma (QGP) at $\theta_I=\pi/2$. 

Locally, the three-dimensional effective theory in the QGP phase is the three-dimensional pure Yang-Mills theory, which is believed to be gapped. 
Thus, we have no massless degrees of freedom in the QGP phase, and the anomaly must be reproduced in a different way from the above two phases.
The critical point is that the degenerated vacua belong to different symmetry-protected topological (SPT) phases protected by
the $\U(1)_{\rml,3} \times \U(1)_\rmb $ symmetry. 
Under the presence of background gauge fields, the effective actions for the two vacua $S_\mathrm{QGP1}$ and $S_\mathrm{QGP2}$ differ by the factor:
\begin{align}
S_\mathrm{QGP1} - S_\mathrm{QGP2}
= -\frac{\im}{2\pi} \int_3 A_{\rml,3} \wedge \diff A_\rmb.
\end{align}
Since the $(\Z_2)_{\mathrm{center}}$ transformation exchanges two vacua, we can reproduce the anomaly as follows:
\begin{align}
S_\mathrm{QGP1} 
\ra
S_\mathrm{QGP2}
= S_\mathrm{QGP1} + \frac{\im}{2\pi} \int_3 A_{\rml,3} \wedge \diff A_\rmb. 
\end{align}

A physical consequence is that the high-temperature domain wall supports Jackiw--Rebbi gapless excitations. 
At $\mu=0$, this high-temperature domain wall is studied in Ref.~\cite{Nishimura:2019umw}, and it supports massless $2$-flavor Schwinger model, which is equivalent to the $\SU(2)$ level-$1$ WZW conformal field theory at low energies (see also Refs.~\cite{Anber:2018jdf,Anber:2018xek, Armoni:2018bga}). 
This means that the domain walls are charged under the chiral symmetry, which partially explains why we can naturally expect the direct transitions between the deconfined phase and the chiral symmetry breaking phase.

We note that, at the large chemical potential with intermediate temperatures, the lattice simulation predicts the existence of \textit{deconfined BCS phase}~\cite{Iida:2019rah}. 
This phase breaks the center symmetry, $(\mathbb{Z}_2)_{\mathrm{center}}\to 1$, and the baryon-number symmetry, $\U(1)_\rmV/\mathbb{Z}_2\to 1$, simultaneously. 
Although each symmetry breaking is sufficient to match the anomaly, the anomaly matching does not prohibit further symmetry breaking. Therefore, the deconfined BCS phase is also consistent with the anomaly matching condition. 

\section{Similarity to $(2+1)$d Quantum Anti-Ferromagnets}\label{sec:analogy}

In this section, we consider a possible and exciting connection between two-flavor QC$_2$D and quantum spin systems. 
As we have discussed in the previous sections, two-flavor massless QC$_2$D must break some of its symmetries for any temperatures and quark chemical potentials at the isospin RW point. 
This situation is nothing but the persistent order discussed in condensed-matter contexts, and let us try to make their connection as concrete as possible.

The form of the mixed anomaly (\ref{eq: anom}) indeed takes almost the same form of the 't~Hooft anomaly of anti-ferromagnetic spin systems in $(2+1)$ dimensions~\cite{Komargodski:2017dmc}. 
At low-energies, that system can be described by the three-dimensional easy-plane $\mathbb{C}P^1$ model:
\begin{align} \label{eq: epcp1}
\int \diff^3x \Big[ |(\diff - \im b) \phi |^2 &+ r |\phi|^2 
+ \lambda|\phi|^4+ \lambda_\mathrm{EP} ( \phi^\+ \sigma_z \phi ) ^2
\Big],
\end{align}
where we introduced a two-component complex scalar $\phi = (\phi_1, \phi_2)^T$ and a dynamical \textit{noncompact} $\U(1)$ gauge field $b$.\footnote{``Noncompact'' means that we do not perform the path integral over the monopole configurations.
Consequently, the model has the $\U(1)$ magnetic symmetry, which plays an important role in the following discussion.}
We schematically suppress the Maxwell term for $b$ because the gauge coupling flows to the strong coupling limit in three dimensions.
While the first three terms respect the $\SO(3)_\mathrm{spin}$ spin symmetry for the spin vector $\phi^\+ \sigma_a \phi$,
the last term explicitly breaks the spin symmetry down to $\mathrm{O}(2)_\mathrm{spin}$ and is called the easy-plane potential. 
For $\lambda_\mathrm{EP}>0$, the spin vector prefers the $xy$-plane in the spin space,
while the spin vector tends to be aligned along the $z$-axis for $\lambda_\mathrm{EP}<0$.
This model describes the unconventional quantum critical point between the N\'eel and valence bond solid (VBS) phases
in $(2+1)$-dimensional anti-ferromagnets,
which is a representative example of phase transitions beyond the Landau--Ginzburg paradigm~\cite{Senthil1490, PhysRevB.70.144407, Wang:2017txt}.\footnote{To be precise, we have to add a monopole operator to the $n$th power
 to take into account the discreteness of the VBS order parameter
 when we consider the N\'eel-VBS transition 
 on the rectangular ($n=2$), honeycomb ($n=3$), and square lattices ($n=4$)~\cite{PhysRevB.70.144407, Read:1989zz, Read:1990zza}.}
It is also attracting attention as a condensed matter application of the $(2+1)$-dimensional dualities~\cite{Seiberg:2016gmd,Karch:2016sxi,Wang:2017txt} (see Ref.~\cite{Senthil:2018cru} for a review).

To clarify the relation between two-flavor QC$_2$D at the isospin RW point and easy-plane $\mathbb{C}P^1$ model, let us more closely look at the global symmetry of (\ref{eq: epcp1}), which is given by 
\be
\begin{split}
(\mathbb{Z}_2)_{\mathcal{C}}\ltimes &
\left[\mathrm{O}(2)_{\mathrm{spin}}\times \U(1)_\rmm\right] 
\\
&\subset 
(\mathbb{Z}_2)_{\mathcal{C}}\ltimes 
\left[\SO(3)_{\mathrm{spin}}\times \U(1)_\rmm\right]. 
\end{split}
\ee
Here, 
$(\mathbb{Z}_2)_{\mathcal{C}}$ is the charge-conjugation symmetry, 
$\mathrm{O}(2)_{\mathrm{spin}}$ is the remnant spin symmetry, and $\U(1)_\rmm$ is the \textit{magnetic} symmetry.
We note that $\mathrm{O}(2)\simeq \mathbb{Z}_2\ltimes \U(1)$, so the $\mathrm{O}(2)_{\mathrm{spin}}$ symmetry is composed of the continuous $\U(1)$ transformation, 
\begin{align}
\U(1)_{\mathrm{spin}}: \phi_1 \ra \rme^{\im \alpha} \phi_1,\; \;
\phi_2 \ra \phi_2,\; \;
b \ra b,
\end{align}
and the discrete $\mathbb{Z}_2$ transformation,
\be
(\mathbb{Z}_2)_{\mathrm{spin}}: \phi_1 \longleftrightarrow \phi_2,\;\; b\to b. 
\ee
More interesting symmetry of (\ref{eq: epcp1}) is the magnetic symmetry, $\U(1)_\rmm$, whose conserved current is given by $\star J_\rmm = \frac{1}{2 \pi} \diff b$.
This symmetry does not acts on the fields in the Lagrangian, but acts on a monopole operator $\mathcal{M}_b$ for the dynamical gauge field $b$~\cite{Borokhov:2002cg} as\footnote{The VBS order parameter in the antiferromagnets is realized as the monopole operator~\cite{Senthil1490}.}
\begin{align}
\mathcal{M}_b \ra \rme^{\im \beta}\mathcal{M}_b.
\end{align}
This symmetry structure motivates us to make a correspondence between two-flavor QC$_2$D at the isospin RW point and the easy-plane $\mathbb{C}P^1$ model as 
\be
\begin{split}
(\mathbb{Z}_2)_{\mathrm{center}}\ltimes \U(1)_{\rmL,3}\, &\Leftrightarrow\, \mathrm{O}(2)_{\mathrm{spin}},\;\; 
\\
\U(1)_\rmV/\mathbb{Z}_2\, &\Leftrightarrow \, \U(1)_\rmm. 
\end{split}
\ee
So far, we have checked that the group structures are indeed the same.

The vital point here is that not only the group structures but also the 't~Hooft anomaly has the same form under this correspondence. 
To see this, let us introduce the $\U(1)_{\mathrm{spin}}$ gauge field $A_\mathrm{spin}$, and then the kinetic term is replaced as follows:
\be
|(\diff - \im b)\phi|^2\Rightarrow |(\diff -\im b -\im A_{\mathrm{spin}})\phi_1|^2 + |(\diff - \im b)\phi_2|^2. 
\ee
Under the presence of $A_{\mathrm{spin}}$, we should modify the $(\mathbb{Z}_2)_{\mathrm{spin}}$ transformation to keep this kinetic term invariant: 
\be 
\phi_1\longleftrightarrow \phi_2,\; b \to b+ A_{\mathrm{spin}},\; A_\mathrm{spin}\to - A_{\mathrm{spin}}. 
\ee
Let us see how it affects the partition function when we also gauge the $\U(1)_\rmm$ symmetry. 
The minimal coupling of the $\U(1)_\rmm$ gauge field, $A_\rmm$, is given by the topological coupling, 
\be
{\im \over 2\pi}\int A_\rmm \wedge \diff b. 
\label{eq: gauged cp1}
\ee
Because of this topological term, \eqref{eq: gauged cp1},
the action is affected by the $(\mathbb{Z}_2)_{\mathrm{spin}}$ transformation, and acquires the overall phase, 
\begin{align}
\label{eq:anomaly_spin}
\frac{\im}{2 \pi} \int_3 A_\rmm \wedge \diff A_\mathrm{spin}.
\end{align}
This is the signature of the $[\mathbb{Z}_2\ltimes \U(1)]_{\mathrm{spin}}\times \U(1)_\rmm$ anomaly.
We here emphasize that this anomaly is identical with the anomaly (\ref{eq: anom}) of QC$_2$D at the isospin RW point.

\begin{figure*}[t]
\begin{center}
\subfloat[][Easy-axis N\'eel phase]{\includegraphics[width=6cm]{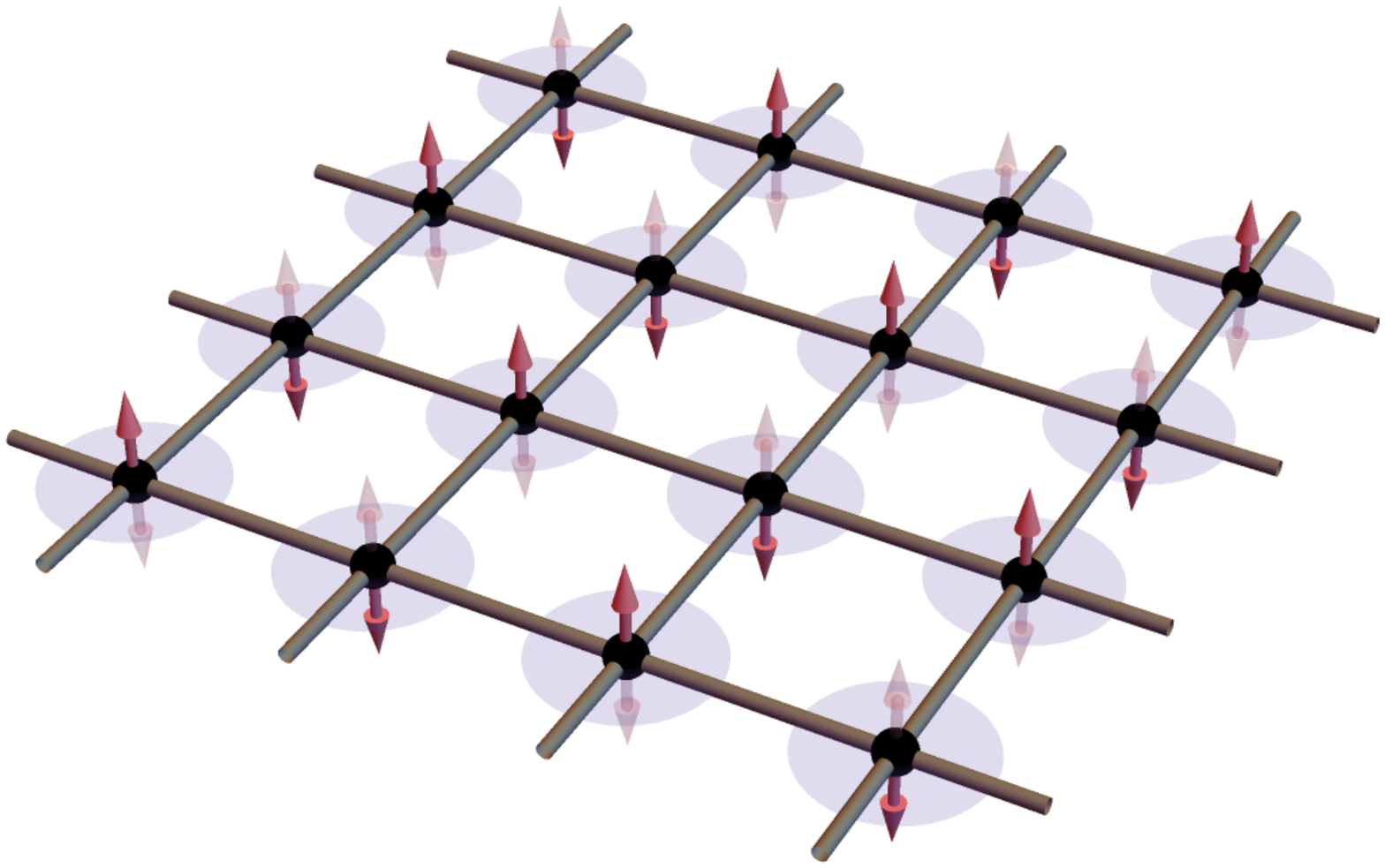}\label{fig:easyaxis}}
\subfloat[][Easy-plane N\'eel phase]{\includegraphics[width=6cm]{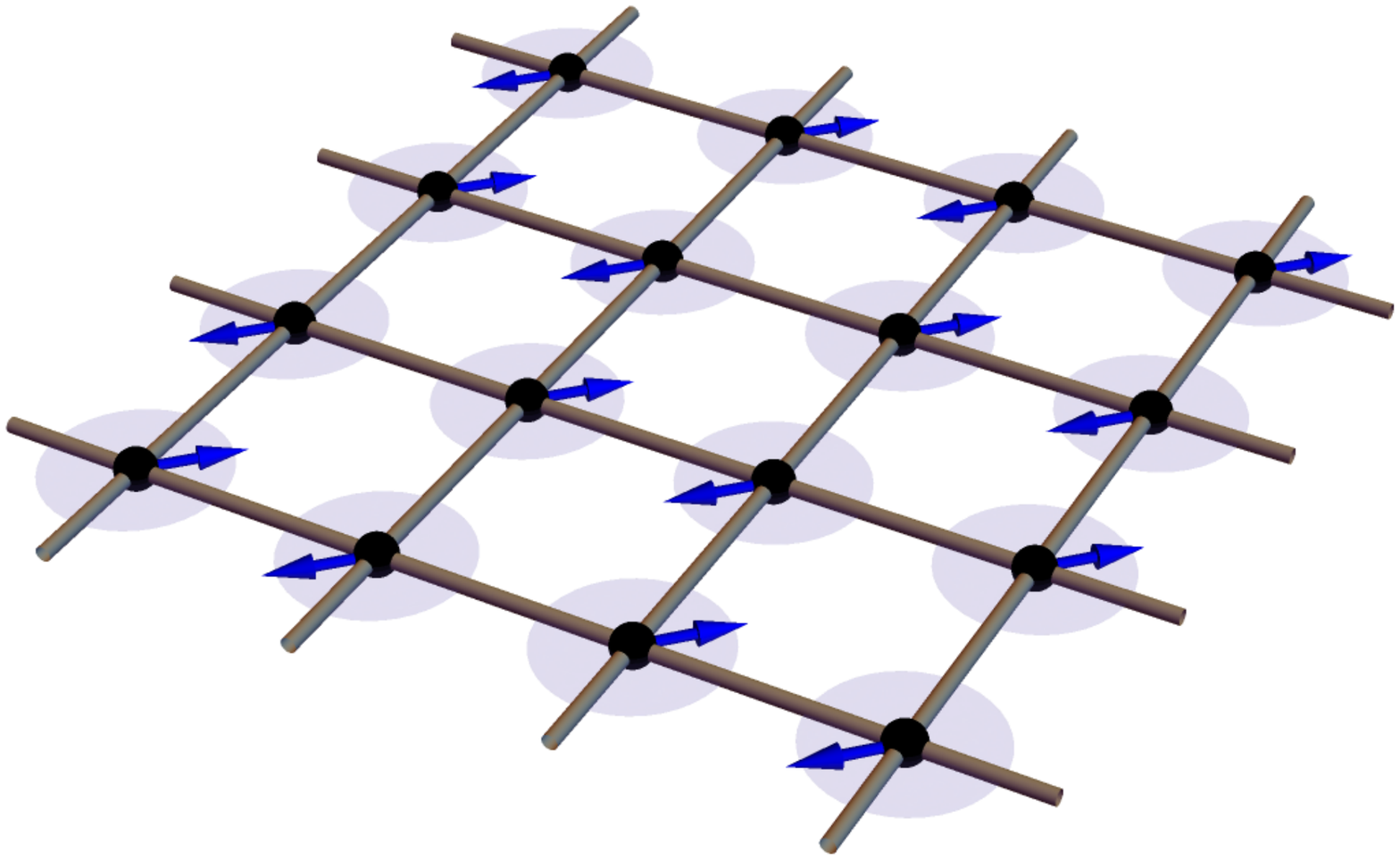}\label{fig:easyplane}}
\subfloat[][VBS phase]{\includegraphics[width=6cm]{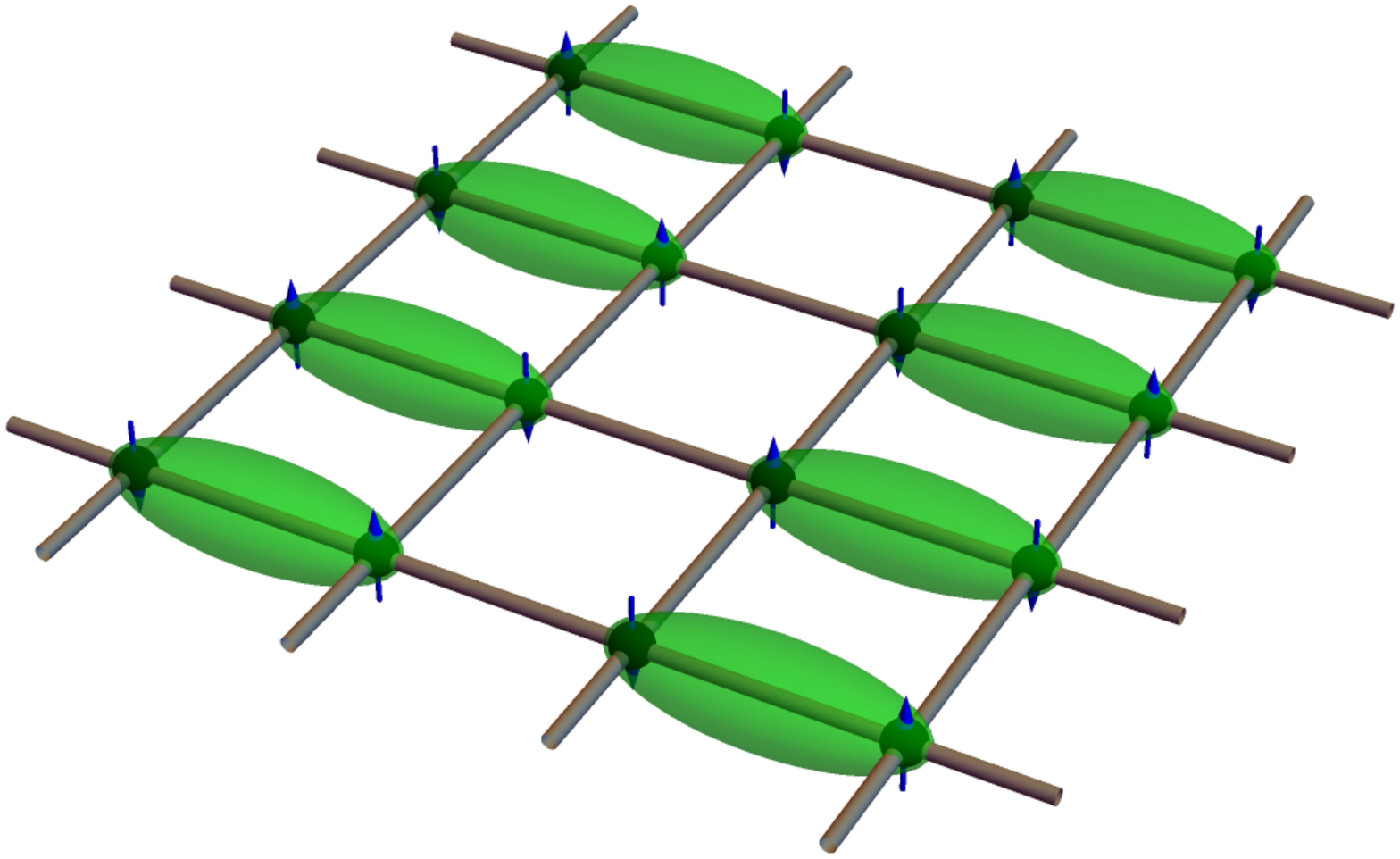}\label{fig:vbs}}
\end{center}
\caption{Illustrations of the phases of the easy-plane $\mathbb{C}P^1$ model.
(Left) In the easy-axis N\'eel phase, the microscopic spins are ordered in a staggered manner and point in the $\pm z$ directions.
(Middle) In the easy-plane N\'eel phase, the spins form an antiferromagnetic order, preferring the $xy$-plane.
(Right) In the valence bond solid phase, the spins are paired with their nearest neighbors to be spin-singlet.
Green ellipses indicate the spin-singlet pairs.
}
\end{figure*}

The identical structures of the global symmetry and 't Hooft anomaly motivate us to translate the QGP, chiral symmetry breaking, and baryon superfluid phases in QC$_2$D in terms of the $\mathbb{C}P^1$ model.
The corresponding phases of the $\mathbb{C}P^1$ model are accessible by tuning the parameters $r$ and $\lambda_\mathrm{EP}$.
Varying $r$, we encounter a phase transition at $r = r_\mathrm{c}$ which separates N\'eel phases ($r<r_\mathrm{c}$) and the VBS phase ($r>r_\mathrm{c}$).
For $r < r_\mathrm{c}$, the complex scalar condenses so that the spin vector $\phi^\+ \sigma_a \phi$ acquires a nonzero expectation value and breaks the spin symmetry.
When $\lambda_\mathrm{EP} = 0$, we obtain the $\SO(3)_\mathrm{spin}$-symmetric N\'eel phase characterized by the symmetry breaking pattern $\SO(3)_\mathrm{spin} \ra \SO(2)_\mathrm{spin}$.
Nevertheless, turning on the easy-plane potential alters the symmetry-breaking pattern in the N\'eel phase.
$(\Z_2)_\mathrm{spin} \subset \mathrm{O}(2)_\mathrm{spin}$ is broken for a negative $\lambda_\mathrm{EP}$ because the spin vector points $\pm z$-direction to reduce the ground state energy.
This phase is known as the easy-axis N\'eel phase, which corresponds to the QGP phase in two-flavor QC$_2$D.
On the other hand, a positive $\lambda_\mathrm{EP}$ confines the spin vector in the $xy$-plane in the spin space, breaking $\U(1)_\mathrm{spin}$ spontaneously.
This phase is called the easy-plane N\'eel phase and the remnant of the chiral symmetry breaking phase in two-flavor QC$_2$D.
We illustrate these phases in Figs.~\ref{fig:easyaxis} and~\ref{fig:easyplane}.

On the other hand, for $r>r_\mathrm{c}$, the complex scalar is gapped and is integrated out.
The effective theory in this phase is thus the three-dimensional Maxwell theory, where the $\U(1)_\mathrm{M}$ is spontaneously broken. 
A photon is understood as a Nambu--Goldstone boson associated with this symmetry breaking.
Therefore, the VBS phase is regarded as baryon superfluid in the QC$_2$D language.
We also comment on a microscopic realization of the VBS phase quickly.
In terms of quantum magnets, the microscopic $1/2$ spins are paired with their nearest neighbors to be spin-singlet
and break the $\Z_4$ lattice rotational symmetry, as shown in Fig.~\ref{fig:vbs}.
Note that the $\Z_4$ anisotropy is expected to be dangerously irrelevant around the critical point~\cite{Senthil1490, PhysRevB.70.144407},
and the rotational symmetry is enlarged to $\U(1)$, which is regarded as $\U(1)_\mathrm{M}$ in the $\mathbb{C}P^1$ model.
In Table~\ref{table:correspondence_phases}, we make the correspondence of phases between these models according to this discussion.

\begin{table}[b]
 \centering
 \begin{tabular}{c  c}
\hline \hline 
QC$_2$D at isospin RW point & Easy-plane $\mathbb{C}P^1$ model
\\ \hline
Quark-gluon plasma, $P$ & Easy-axis N\'eel, $ \phi^\+ \sigma_z \phi$
\\
Chiral symmetry breaking, $\sigma + \im \pi_0$  &  Easy-plane N\'eel, $ \phi^*_2 \phi_1$ 
\\
Baryon superfluid, $\Delta$ &Valence bond solid, $\mathcal{M}_b$
\\
\hline \hline
\end{tabular}
 \caption{ Correspondence of the phases and order parameters in two-flavor QC$_2$D and the easy-plane $\mathbb{C}P^1$ model. } 
\label{table:correspondence_phases}
\end{table}

We note that the charge-conjugation symmetry does not enter the anomaly. 
However, since it has the structure of the semi-direct product, the correspondence of symmetries between these models looks to be slightly different when we combine it with $(\mathbb{Z}_2)_{\mathrm{spin}}$ symmetry. 
Let us denote this combined symmetry as $(\mathbb{Z}_2)_{\mathcal{C}+\mathrm{spin}}$, and then the symmetry looks as 
\be
[(\mathbb{Z}_2)_{\mathcal{C}+\mathrm{spin}}\ltimes \U(1)_\rmm] \times \U(1)_{\mathrm{spin}}. 
\ee
Therefore, the roles of $\U(1)_\rmm$ and $\U(1)_{\mathrm{spin}}$ may be interchanged depending on which $\mathbb{Z}_2$ symmetry is chosen for the correspondence. 
This $(\mathbb{Z}_2)_{\mathcal{C}+\mathrm{spin}}$ transformation acts on the dynamical and background fields as 
\be
 \phi_1\longleftrightarrow \phi_2^*,\; b\to -b-A_{\mathrm{spin}},\; A_{\mathrm{spin}}\to A_{\mathrm{spin}},\; A_\rmm\to -A_\rmm. 
\ee
We can indeed check that we obtain the same anomaly (\ref{eq:anomaly_spin}). 
Therefore, the easy-plane N\'eel and VBS phases in Table~\ref{table:correspondence_phases} should also be exchanged when we respect $(\mathbb{Z}_2)_{\mathcal{C}+\mathrm{spin}}$.  
In view of anomaly, both choices are equally good to make up the correspondence.

Here, we have clarified that QC$_2$D at the isospin RW point has a very similar structure with easy-plane $\mathbb{C}P^1$ model. 
They share not only the same symmetry group, but also the same 't~Hooft anomaly, which is highly nontrivial. 
We are tempted to ask if their similarity extends to the dynamics, i.e., the nature of phase transitions. 
The $(2+1)$d easy-plane $\mathbb{C}P^1$ model is now providing a typical example of quantum phase transitions, called deconfined quantum criticality. 
The phase transition point between the easy-plane N\'eel and VBS phases is believed to acquire the emergent symmetry $\SO(4) \supset \U(1)_\mathrm{spin} \times \U(1)_\mathrm{M}$, which can rotate the N\'eel and VBS order parameters.
Furthermore, the emergent symmetry is enlarged to $\SO(5) \supset (\Z_2)_\mathrm{spin} \times \SO(4)$ when the $\SO(3)_\mathrm{spin}$ symmetry recovers (i.e., $\lambda_\mathrm{EP} = 0$)\footnote{See Refs.~\cite{Tanaka:2005imi,Senthil:2005jk} for theoretical studies, Ref.~\cite{Nahum:2015vka} for numerical observation, and Ref.~\cite{Wang:2017txt} for recent theoretical development.}. 
Recall that the $\SO(4)$ symmetry mixing $\U(1)_{\rmL.3}$ and $\U(1)_{\rmV}/\Z_2$ does appear in chiral symmetry breaking phase at $\mu=0$.
Thus, this scenario may indicate an emergence of an intriguing $\SO(5)$ symmetry, that rotates the $\SO(4)$ chiral condensate ($\sigma$, $\pi$, $\Delta_1$, and $\Delta_2$) and the Polyakov loop $P$.
Therefore, it sounds like a very reasonable question in this context if the similar enhancement of symmetry occurs inside the phase diagram of QC$_2$D. 
We note that this depends on the dynamics of QC$_2$D, so we must go beyond the kinematic analysis based on symmetry and anomalies to answer this question. We shall leave it for possible future studies.

\section{Summary}
\label{sec:summary}

In this paper, we have studied the phase diagram of two-flavor massless QC$_2$D at the isospin RW point based on the 't~Hooft anomaly matching. 
We note that this setup does not suffer from the sign problem, because the Dirac determinant over the $u$-quark sector is related to that of $d$-quark sector by complex conjugation. 
Therefore, we can compare our study with numerical lattice simulations at finite quark chemical potentials with imaginary isospin chemical potentials. 

We first gave a careful review of the global symmetry of massless QC$_2$D, especially paying attention to its global nature, such as the discrete symmetry and possible quotients by discrete factors. 
For the computation of perturbative anomalies, such details are not essential. 
On the other hand, the careful treatment of the discrete parts becomes crucial when we discuss the more subtle global anomaly. 
Especially, when the imaginary isospin chemical potential takes the special value, $\theta_I=-\im \mu_I L=\pi/2$, two-flavor QC$_2$D enjoys the $(\mathbb{Z}_2)_\mathrm{center}$ symmetry, which acts on both the quark flavors and the Polyakov loop. 
Because of this fact, we concentrate on studying the phase diagram at the isospin RW point, $\theta_I=\pi/2$ in this paper. 

After studying the perturbative anomaly to get an insight on $T=0$, we computed the discrete mixed 't~Hooft anomaly, which involves the center symmetry $(\mathbb{Z}_2)_{\mathrm{center}}$, the baryon-number symmetry $\U(1)_\rmV/\mathbb{Z}_2$, and the isospin chiral symmetry, $\U(1)_{\rmL,3}$.
This is the main result of this paper. 
This anomaly provides a meaningful constraint on the phase diagram of massless two-flavor QC$_2$D.
In order to satisfy the anomaly matching condition, the phase diagram of massless QC$_2$D at $\theta_I=\pi/2$ basically has to break one of the above three symmetries spontaneously at any temperatures $T$ and quark chemical potentials $\mu$. 
Combined with the study of chiral effective Lagrangian and the numerical results of lattice simulations, we argue that this constraint is indeed satisfied, and we comment on how each phase matches this anomaly. 

Interestingly, the discrete anomaly for massless QC$_2$D at the isospin RW point is very similar to that of $(2+1)$d quantum anti-ferromagnetic systems. 
We propose an explicit correspondence between these two systems based on the consideration on their symmetries and anomalies. 
It would be an interesting future study to consider if this kinematic similarity extends to the similarity of dynamics between these theories. 

\acknowledgments
The authors thank Y. Nishida for useful discussions.
T.F. was supported by RIKEN Junior Research Associate Program when we started this work
and by JSPS KAKENHI Grant Number JP20J13415 after April.
E.I. was supported by the HPCI-JHPCN System Research Project (Project ID: jh200031)
and by Grants-in-Aid for Scientific Research through Grant No.19K03875
, which were provided by the Japan Society for the Promotion of Science (JSPS).
Discussions during the workshop ``${\mathbb{C}}$P$^{N}$ models: recent development and future directions'' held at Keio university were useful to initiate this work.

\appendix

\section{Convention of Gamma Matrices and Spinors}\label{sec:gamma_matrix}

We here summarize the convention of Euclidean gamma matrices in  this paper. The flat space metric is $g_{\mu\nu}=\delta_{\mu\nu}=\mathrm{diag}(+1,+1,+1,+1)$. 
Weyl representation of the Euclidean gamma matrices is 
\be
\gamma^\mu = \begin{pmatrix}
0 & \sigma^{\mu}\\
\overline{\sigma}^{\mu} & 0
\end{pmatrix}, 
\ee
where $\sigma^{\mu}=(\bm{1}, \im \sigma_1, \im \sigma_2, \im \sigma_3)$ and $\overline{\sigma}^\mu=(\sigma^\mu)^\dagger=(\bm{1}, -\im \sigma_1, -\im \sigma_2, -\im \sigma_3)$. 
In the dotted and undotted spinor notation, spin indices are assigned as $(\sigma^\mu)_{\alpha\dot{\beta}}$ and $(\overline{\sigma}^\mu)^{\dot{\alpha}\beta}$, where $\alpha, \dot{\alpha},\ldots\in \{1,2\}$. 
We describe the undotted spinor as $\psi_\alpha$ and the dotted spinor as $\bar{\tilde{\psi}}_{\dot{\beta}}$, and their conjugate fields are $\bar{\psi}_{\dot{\alpha}}$ and $\tilde{\psi}_\beta$, respectively\footnote{Let us comment on the physical interpretation about the symbols of spinors. Both $\psi_\alpha$ and $\tilde{\psi}_\alpha$ denote left-handed Weyl fermions under the Lorentz transformation, i.e. in the $(\bm{2},\bm{1})$ representation of $\mathrm{Spin}(4)\simeq \SU(2)\times \SU(2)$. In the context of QCD, however, we usually interpret $\psi_\alpha$ as the left-handed quark ($=\psi_{\mathrm{D},\rmL}$), and $\ve^{\alpha\beta}\tilde{\psi}_\beta$ as the anti-particle of right-handed quark ($=\overline{\psi}_{\mathrm{D},\rmR}\,$). }.
These are Weyl fermions, and the Dirac fermions are defined as\footnote{When we use the Dirac spinor, we always put the subscript ``$\mathrm{D}$'' throughout the paper. }
\be
\psi_\mathrm{D}=\begin{pmatrix}
\psi_\alpha\\
\ve^{\dot{\alpha}\dot{\beta}}\bar{\tilde{\psi}}_{\dot{\beta}}
\end{pmatrix},\quad 
\overline{\psi}_\mathrm{D}=\begin{pmatrix}
\ve^{\alpha\beta}\tilde{\psi}_\beta& \bar{\psi}_{\dot{\alpha}}
\end{pmatrix}. 
\ee
In this convention, the Dirac Lagrangian can be written as 
\be
\overline{\psi}_\mathrm{D}\gamma^\mu \p_\mu \psi_\mathrm{D}=\bar{\psi}_{\dot{\alpha}}(\overline{\sigma}^\mu)^{\dot{\alpha}\alpha}\p_\mu \psi_\alpha+\bar{\tilde{\psi}}_{\dot{\beta}}(\overline{\sigma}^\mu)^{\dot{\beta}\beta}\p_\mu \tilde{\psi}_\beta, 
\ee
up to the integration by parts. The Dirac mass, or chiral condensate, can be written as 
\be
\overline{\psi}_\mathrm{D}\psi_\mathrm{D}=\ve^{\alpha\beta}\tilde{\psi}_\beta \psi_\alpha+\ve^{\dot{\alpha}\dot{\beta}}\bar{\psi}_{\dot{\alpha}}\bar{\tilde{\psi}}_{\dot{\beta}}. 
\ee 
Below, we omit spin indices and also the $\ve$ tensor for simplicity, when the way of contraction is evident. 

When we consider the vector-like $\SU(N_c)$ gauge theory, $\psi$ belongs to the defining representation, $\bm{N}_c$, and $\tilde{\psi}$ belongs to its conjugate representation, $\overline{\bm{N}}_c$, so that $\psi_D$ transforms as $\bm{N}_c$. The Dirac Lagrangian with minimal coupling is given as  
\be
\overline{\psi}_\mathrm{D} \gamma^\mu (\p_\mu+\im a_\mu)\psi_\mathrm{D}= \bar{\psi} \overline{\sigma}^\mu(\p_\mu+\im a_\mu) \psi+\bar{\tilde{\psi}} \overline{\sigma}^\mu (\p_\mu-\im a_\mu)\tilde{\psi}. 
\ee 
When we consider $N_f$ Dirac flavors, the list of charges under $\SU(N_c)$ gauge symmetry and global chiral symmetry, $[\SU(N_f)_\rmL\times \SU(N_f)_\rmR\times \U(1)_\rmV]/\mathbb{Z}_{N_f}$, can be summarized as follows: 
\be
\begin{array}{c||c||c|c|c}
	& \SU(N_c) & \SU(N_f)_\rmL & \SU(N_f)_\rmR & \U(1)_\rmV \\ \hline
\psi & \bm{N}_c & \bm{N}_f & \bm{1} & 1\\
\tilde{\psi} & \overline{\bm{N}}_c & \bm{1} & \overline{\bm{N}}_f & -1
\end{array}
\label{eq:charge_table_quarks_SU(Nc)}
\ee
We note that this fits into the standard convention in supersymmetric QCD. 

When $N_c=2$, the defining representation can be identified with its conjugate representation, $\overline{\bm{2}}\simeq \bm{2}$, by pseudo-reality of $\SU(2)$. Because of this, it is more convenient to take $\tilde{\psi}$ in the defining representation, and the Dirac fermion can be represented as 
\be
\psi_\mathrm{D}=\begin{pmatrix}
\psi\\
(\ve_{\mathrm{color}}\otimes \ve_{\mathrm{spin}}) \bar{\tilde{\psi}}
\end{pmatrix}, 
\label{eq:Dirac_Weyl_appendix}
\ee
where $\ve_{\mathrm{color}}(=\im\tau_2)$ is the invariant tensor of the $\SU(2)$ color space, and we also denote the invariant tensor of $\mathrm{Spin}(4)\simeq \SU(2)\times \SU(2)$ as $\ve_{\mathrm{spin}}$ in order to avoid confusions. 
Throughout the main text of this paper, we use this convention for Weyl and Dirac spinors. 
We readily find that the Dirac Lagrangian becomes 
\be
\overline{\psi}_\mathrm{D} \gamma^\mu (\p_\mu+\im a_\mu)\psi_\mathrm{D}= \bar{\psi} \overline{\sigma}^\mu(\p_\mu+\im a_\mu) \psi+\bar{\tilde{\psi}} \overline{\sigma}^\mu (\p_\mu+\im a_\mu)\tilde{\psi}, 
\ee
because $(\ve_{\mathrm{color}})^T a^T (\ve_{\mathrm{color}})=-a$ for $\SU(2)$ gauge field $a$. 
When we consider $N_f$ Dirac flavors, the charge table (\ref{eq:charge_table_quarks_SU(Nc)}) is modified as 
\be
\begin{array}{c||c||c|c|c}
	& \SU(2) & \SU(N_f)_\rmL & \SU(N_f)_\rmR & \U(1)_\rmV \\ \hline
\psi & \bm{2} & \bm{N}_f & \bm{1} & 1\\
\tilde{\psi} & \bm{2} & \bm{1} & \overline{\bm{N}}_f & -1
\end{array}
\label{eq:charge_table_appendix}
\ee
As we explain in Sec.~\ref{sec:symmetry}, we can now rotate $\psi$ and $\tilde{\psi}$ as a global symmetry because they share the same color and Lorentz structures, so it is now easy to see that the chiral symmetry is extended as $\SU(2N_f)\supset [\SU(N_f)_\rmL\times \SU(N_f)_\rmR\times \U(1)_\rmV]/\mathbb{Z}_{N_f}$~\cite{Smilga:1994tb, Peskin:1980gc}. 

\section{On Symplectic Group, $\Sp(N)$, and Vacuum Manifold $\SU(2N)/\Sp(N)$}\label{sec:symplectic}

We here summarize properties of compact symplectic group, $\Sp(N)$, and its relation to $\SU(2N)$~\cite{Peskin:1980gc}, which are important to understand chiral symmetry breaking of QC$_2$D. 

Let us start with the definition of $\Sp(N)$. It is convenient to introduce the non-compact symplectic group $\Sp(2N,\mathbb{C})$, defined by  
\be
\Sp(2N,\mathbb{C})=\{M\in \mathrm{GL}(2N, \mathbb{C}) \,|\, M^T \Omega_N M=\Omega_N\},
\ee
where 
\be
\Omega_N=\begin{pmatrix}
\bm{0} & \bm{1}_N\\
-\bm{1}_N & \bm{0}
\end{pmatrix}
\ee
is the symplectic form on $\mathbb{C}^{2N}$. 
We can show that $\mathrm{det}(M)=1$ from the condition $M^T\Omega_N M=\Omega_N$, and thus $\Sp(2N,\mathbb{C})\subset \mathrm{SL}(2N,\mathbb{C})$. This is a simple Lie group, which is noncompact and simply-connected. 
Now, compact symplectic group, $\Sp(N)$, is defined by 
\be
\Sp(N) = \Sp(2N,\mathbb{C})\cap \SU(2N). 
\ee
That is, if and only if $U\in \SU(2N)$ satisfies $U^T \Omega_N U=\Omega_N$, $U\in \Sp(N)$. 
Since unitary matrices satisfy $U^T=(U^*)^{-1}$, the condition $U\Omega_N U^T=\Omega_N$ is also equivalent by taking complex conjugation of $(U^{-1})^T \Omega_N U^{-1}=\Omega_N$. 
In other words, $U\in \Sp(N) \Leftrightarrow U^*\in \Sp(N)$ for $U\in \SU(2N)$. 

An important property of $\Sp(N)$ as a subgroup of $\SU(2N)$ comes out by considering the anti-symmetric two-index representation of $\SU(2N)$. 
More concretely, we define a $(2N\times 2N)$ matrix-valued bosonic field, $\Sigma_{ij}$, using the fermionic field $\Psi_i$ in the defining representation of $\SU(2N)$ as 
\be
\Sigma_{ij}=\Psi_i\Psi_j. 
\ee 
By the anti-commutativity of fermions, $\Sigma^T=-\Sigma$, and this belongs to the two-index anti-symmetric representation. 
Under the $\SU(2N)$ transformation, $U\to U\Psi$, 
\be
\Sigma\mapsto U\Sigma U^T. 
\ee
Let us pick up a specific point, 
\be
\Sigma=\Sigma_0\equiv \Omega_N.
\ee 
Then, the stabilizer subgroup of $\Sigma_0$ in $\SU(2N)$ is given by 
\be
\{U\in \SU(2N) \,|\, U \Sigma_0 U^T=\Sigma_0\}=\Sp(N). 
\ee
As a consequence, when the bosonic field $\Sigma$ condenses as $\Sigma=\Sigma_0$, the spontaneous breaking pattern is $\SU(2N)\to \Sp(N)$. 
The vacuum manifold of this SSB is given by the symmetric space
\be
\SU(2N)/\Sp(N)\simeq \{U\Sigma_0 U^T \, |\, U\in \SU(2N) \}, 
\ee
which is a connected subspace of $(2N\times 2N)$ anti-symmetric matrices with determinant $1$. 
Its dimension is $2N^2-N-1$. 

Since we especially pay attention to the $2$-flavor case in this paper, it would be useful to closely look at the case $N=2$. 
In this case, it is convenient to use the exceptional isomorphisms, 
\be
\mathrm{Spin}(6)\simeq \SU(4),\; \mathrm{Spin}(5)\simeq \Sp(2). 
\ee
Since $\Sigma$ is in a two-index representation of $\SU(4)$, we can regard it as in a representation of $\SU(4)/\mathbb{Z}_2\simeq \mathrm{Spin}(6)/\mathbb{Z}_2\simeq  \SO(6)$. 
Indeed, the two-index anti-symmetric representation of $\SU(4)$ is nothing but the defining representation of $\SO(6)$, i.e. $\Sigma\in \mathbb{R}^6$. 
If $\Sigma_0\not =0$, this means the spontaneous breaking
\be
\SO(6)\simeq{\SU(4)\over \mathbb{Z}_2}\to \SO(5)\simeq {\Sp(2)\over \mathbb{Z}_2},
\ee
which fits the general argument by putting $N=2$. The vacuum manifold is given by $\SO(6)/\SO(5)\simeq S^5$.

\bibliographystyle{utphys}
\bibliography{
2-color-QCD,
QFT}

\end{document}